\documentclass[a4paper,UKenglish]{lipics}

\usepackage[]{microtype}
\usepackage[utf8]{inputenc}
\usepackage[]{proof}
\usepackage[]{latexsym}
\usepackage[]{eqparbox}
\usepackage[]{amssymb}
\usepackage[]{amsmath}

\let\oldampersand\&

\renewcommand*\&{{\itshape\oldampersand}}

\title{The Rooster and the Syntactic Bracket\footnote{This research has received funding from the European Research Council under the FP7 grant agreement 278673, Project MemCAD}}

\titlerunning{The Rooster and the Syntactic Bracket}

\author[1]{Hugo Herbelin}
\author[2]{Arnaud Spiwack}

\affil[1]{Inria Paris-Rocquencourt\\
Paris, France\\
\texttt{hugo.herbelin@inria.fr}}
\affil[2]{Inria Paris-Rocquencourt\\
Paris, France\\
\texttt{arnaud@spiwack.net}}

\authorrunning{H. Herbelin and A. Spiwack}

\Copyright{Hugo Herbelin and Arnaud Spiwack}

\subjclass{F.3.3 Studies of Program Constructs}

\keywords{Coq, Calculus of inductive constructions, Impredicativity, Strictly positive type families, Inductive type families}

\serieslogo{}
\volumeinfo
  {Billy Editor, Bill Editors}
  {2}
  {Conference title on which this volume is based on}
  {1}
  {1}
  {1}
\EventShortName{}
\DOI{10.4230/LIPIcs.xxx.yyy.p}

\begin{document}
  \maketitle
  \begin{abstract}
      We propose an extension of pure type systems with an algebraic presentation of inductive and co-inductive type families with proper indices. This type theory supports coercions toward from smaller sorts to bigger sorts via explicit type construction, as well as impredicative sorts. Type families in impredicative sorts are constructed with a bracketing operation. The necessary restrictions of pattern-matching from impredicative sorts to types are confined to the bracketing construct. This type theory gives an alternative presentation to the calculus of inductive constructions on which the Coq proof assistant is an implementation.
  \end{abstract}
  \section{Introduction}
  \par
  In the Coq proof assistant~\cite{coq} inductive types are treated as toplevel definitions. If it makes sense from a convenience or an efficiency point of view, the monolithic nature of the definitions make it hard to describe what they precisely mean. As a matter of fact, inductive definitions mean different things depending on the type they are defined in: specifically, some types are interpreted differently in impredicative sorts like \textbf{\textsf{Prop}} or the impredicative variant of \textbf{\textsf{Set}}.
  \par
  In this article, we present a calculus of inductive and co-inductive
  constructions where inductive and co-inductive types are presented
  algebraically. The algebraic presentation is an extension of a PTS~\cite{Barendregt}
  with inductive and co-inductive type families. Thanks to its
  modularity, it is meant to serve as a description which is simpler to expose and more
  mathematically amenable than the monolithic scheme which
  is found in a practical system such as Coq. For the sake of clarity,
  the system is given with a single universe and explicit subtyping,
  although Coq has an unbounded cumulative hierarchy of universes and
  implicit subtyping. Apart from these technicalities, it is believed that our calculus of
  algebraic inductive and co-inductive constructions expresses all
  the features of the Set-impredicative Calculus of Inductive Constructions
  that Coq implements, \emph{e.g.} in its version 8.4 when launched with option
  \symbol{45}impredicative\symbol{45}set.
  \par
  This work draws most of its inspiration from Morris \& al~\cite{Morris2007a,Morris2009} for the algebraic presentation of inductive type families in a predicative sort, and Awodey \& Bauer~\cite{Awodey2004} for the treatment of impredicative sorts.
  \par
  We use examples from Coq to illustrate the algebraic presentation. To differentiate expressions in Coq from expressions in the algebraic presentation, the former are typeset in a sans-serif font and the latter in a roman font.
  \section{Pure type systems}
  \par
  To model the type system of Coq, we start with the classic presentation of pure type systems (\textsc{pts}) of Barendregt~\cite{Barendregt}, which we will then extend to model type families. A \textsc{pts} is characterised by a \emph{single} syntactic category of terms which are used both as ${\lambda}$-terms and as types. It has a single form of typing judgment $\Gamma \vdash u : A$, where $u$ and $A$ are terms, and $\Gamma $ a context assigning terms to variables. A \textsc{pts} has a set of \emph{sorts}, which we shall denote schematically by the symbol $s$. Every sort is an atomic term. A \textsc{pts} has a conversion relation $u\equiv v$. Here we diverge from the presentation of~\cite{Barendregt} which always uses ${\beta}$-conversion. Coq, on the other hand, uses ${\beta}$${\eta}$-conversion on the fragment described in this section. In this article we will take the conversion rule as abstract, not even requiring it to be decidable. We will only require that it contains all the reduction rules which are given in the form $u\leadsto v$ (in this section, we only have ${\beta}$-reduction).
  \par
  \begin{figure}[tp]
    \centering{}\makebox[1.\textwidth]{\parbox{0.5\textwidth}{
    \begin{center}
      $\infer{\Gamma , x{:}A\vdash x : A}{\Gamma \vdash A : s & x\mbox{\textrm{ is fresh in }}\Gamma }$
    \end{center}
    }\parbox{0.5\textwidth}{
    \begin{center}
      $\infer{\Gamma , x{:}B\vdash u : A}{\Gamma \vdash u : A & \Gamma \vdash B : s & x\mbox{\textrm{ is fresh in }}\Gamma }$
    \end{center}
    }}\\
    \makebox[1.\textwidth]{\parbox{0.5\textwidth}{
    \begin{center}
      $\infer{\Gamma \vdash \lambda {x}^{A}.\,u : \prod _{x{:}A}B}{\Gamma \vdash \prod _{x{:}A}B : s & \Gamma , x{:}A\vdash u : B}$
    \end{center}
    }\parbox{0.5\textwidth}{
    \begin{center}
      $\infer{\Gamma \vdash {u}\,{v} : {B}{\left[ \,x\,{\setminus }\,v\,\right] }}{\Gamma \vdash u : \prod _{x{:}A}B & \Gamma \vdash v : A}$
    \end{center}
    }}\\
    \makebox[1.\textwidth]{\parbox{0.5\textwidth}{
    \begin{center}
      $\infer{\Gamma \vdash u : B}{\Gamma \vdash u : A & \Gamma \vdash B : s & A\equiv B}$
    \end{center}
    }\parbox{0.5\textwidth}{
    \begin{center}
      ${\left( \lambda {x}^{A}.\,u\right) }\,{v}\leadsto {u}{\left[ \,x\,{\setminus }\,v\,\right] }$
    \end{center}
    }}\\
    \caption{Generic rules of \textsc{pts}}\label{latex_lib_label_1}
  \end{figure}
  The typing rules of a \textsc{pts} comprise of a set of generic rules given in Figure~\ref{latex_lib_label_1}, together with a number of rules of the form
  \begin{displaymath}
    \parbox{1.\textwidth}{
    \begin{center}
      $\infer{\vdash {s}_{1} : {s}_{2}}{}$
    \end{center}
    }
  \end{displaymath}
  called axioms, and rules of the form
  \begin{displaymath}
    \parbox{1.\textwidth}{
    \begin{center}
      $\infer{\Gamma \vdash \prod _{x{:}A}B : {s}_{3}}{\Gamma \vdash A : {s}_{1} & \Gamma , x{:}A\vdash B : {s}_{2}}$
    \end{center}
    }
  \end{displaymath}
  of product formation rules. As usual we write $A\rightarrow B$ for $\prod _{x{:}A}B$ when $x$ does not bind a variable in $B$.
  \par
  As a starting point of the algebraic presentation, we shall use a \textsc{pts} with two sorts, $\mathrm{Type}$ and $\Box $, together with the following axiom:
  \begin{displaymath}
    \parbox{1.\textwidth}{
    \begin{center}
      $\infer{\Gamma \vdash \mathrm{Type} : \Box }{}$
    \end{center}
    }
  \end{displaymath}
  and the following product formation rule:
  \begin{displaymath}
    \parbox{1.\textwidth}{
    \begin{center}
      $\infer{\Gamma \vdash \prod _{x{:}A}B : {\mathrm{max}}\,{{s}_{1}}\,{{s}_{2}}}{\Gamma \vdash A : {s}_{1} & \Gamma , x{:}A\vdash B : {s}_{2}}$
    \end{center}
    }
  \end{displaymath}
  where ${\mathrm{max}}\,{s}\,{s}=s$ and ${\mathrm{max}}\,{\mathrm{Type}}\,{\Box }={\mathrm{max}}\,{\Box }\,{\mathrm{Type}}=\Box $
  \par
  The sorts $\mathrm{Type}$ and $\Box $ are predicative. The sort $\Box $ plays a technical role in allowing type variable and the formation of type-level functions; it cannot, however, be referenced in terms. In the following sections, $\Box $ will also be used to be able to define types by pattern-matching (\emph{strong elimination}).
  \par
  To model the entire Coq system, $\mathrm{Type}$ and $\Box $ would be replaced with a hierarchy of predicative sorts ${\mathrm{Type}}_{i}$, such that
  \begin{displaymath}
    \parbox{1.\textwidth}{
    \begin{center}
      $\infer{\Gamma \vdash {\mathrm{Type}}_{i} : {\mathrm{Type}}_{i+1}}{}$
    \end{center}
    }
  \end{displaymath}
  are axioms. Adapting the presentation to a hierarchy of sorts is straightforward, but in the interest of keeping to the heart of the matter we give a presentation with two sorts.
  \section{Inductive type families}
  We shall now extend the algebraic presentation with a notion of \emph{inductive type families} to model (dependent) datatypes. In this section we will stay in the predicative fragment of Coq.
  \par
  Contrary to the inductive types of Coq, where inductive definitions must be \emph{named} at \emph{toplevel}, like in:
  \begin{displaymath}
    \parbox{0.8\textwidth}{
    \begin{flushleft}
      \textbf{\textsf{Inductive}}\hphantom{ }\textsf{Even}\hphantom{ }\symbol{58}\hphantom{ }\textbf{\textsf{Type}}\hphantom{ }\symbol{58}\symbol{61}\\
      ${\mid}$\hphantom{ }\textsf{eo}\hphantom{ }\symbol{58}\hphantom{ }\textsf{Even}\\
      ${\mid}$\hphantom{ }\textsf{es}\hphantom{ }\symbol{58}\hphantom{ }\textsf{Odd}\hphantom{ }${\rightarrow}$\hphantom{ }\textsf{Even}\\
      \textbf{\textsf{with}}\hphantom{ }\textsf{Odd}\hphantom{ }\symbol{58}\hphantom{ }\textbf{\textsf{Type}}\hphantom{ }\symbol{58}\symbol{61}\\
      ${\mid}$\hphantom{ }\textsf{os}\hphantom{ }\symbol{58}\hphantom{ }\textsf{Even}\hphantom{ }${\rightarrow}$\hphantom{ }\textsf{Odd}\symbol{46}
    \end{flushleft}
    }
  \end{displaymath}
  the presentation given here is essentially anonymous, and inductive families need not be defined at toplevel prior to use. Mutually inductive types such as \textsf{Even} and \textsf{Odd} are not modelled directly, rather they are encoded using an index:
  \begin{displaymath}
    \parbox{0.8\textwidth}{
    \begin{flushleft}
      \textbf{\textsf{Inductive}}\hphantom{ }\textsf{EvenOdd}\hphantom{ }\symbol{58}\hphantom{ }\textsf{bool}\hphantom{ }${\rightarrow}$\hphantom{ }\textbf{\textsf{Type}}\hphantom{ }\symbol{58}\symbol{61}\\
      ${\mid}$\hphantom{ }\textsf{eo}\hphantom{ }\symbol{58}\hphantom{ }\textsf{EvenOdd}\hphantom{ }\textsf{true}\\
      ${\mid}$\hphantom{ }\textsf{es}\hphantom{ }\symbol{58}\hphantom{ }\textsf{EvenOdd}\hphantom{ }\textsf{false}\hphantom{ }${\rightarrow}$\hphantom{ }\textsf{EvenOdd}\hphantom{ }\textsf{true}\\
      ${\mid}$\hphantom{ }\textsf{os}\hphantom{ }\symbol{58}\hphantom{ }\textsf{EvenOdd}\hphantom{ }\textsf{true}\hphantom{ }${\rightarrow}$\hphantom{ }\textsf{EvenOdd}\hphantom{ }\textsf{false}\symbol{46}\\
      \textbf{\textsf{Definition}}\hphantom{ }\textsf{Even}\hphantom{ }\symbol{58}\symbol{61}\hphantom{ }\textsf{EvenOdd}\hphantom{ }\textsf{true}\symbol{46}\\
      \textbf{\textsf{Definition}}\hphantom{ }\textsf{Odd}\hphantom{ }\symbol{58}\symbol{61}\hphantom{ }\textsf{EvenOdd}\hphantom{ }\textsf{false}\symbol{46}
    \end{flushleft}
    }
  \end{displaymath}
  \par
  This encoding works as long as all the mutual definitions are all in the same sort. A variant for mutual definition involving \textbf{\textsf{Type}} and \textbf{\textsf{Prop}} is demonstrated in Section~\ref{latex_lib_label_8}. When the types being defined are in different predicative sorts, however, we have to resort to another encoding which involves nested datatypes~\cite[Section 8.6]{barras2013}.
  \par
  We will not explore the latter kind of mutual definition. However, nested datatypes -- where recursive calls occur as arguments of another type -- such as:
  \begin{displaymath}
    \parbox{0.8\textwidth}{
    \begin{flushleft}
      \textbf{\textsf{Inductive}}\hphantom{ }\textsf{List}\hphantom{ }\symbol{40}\textsf{A}\symbol{58}\textbf{\textsf{Type}}\symbol{41}\hphantom{ }\symbol{58}\hphantom{ }\textbf{\textsf{Type}}\hphantom{ }\symbol{58}\symbol{61}\\
      ${\mid}$\hphantom{ }\textsf{nil}\hphantom{ }\symbol{58}\hphantom{ }\textsf{List}\hphantom{ }\textsf{A}\\
      ${\mid}$\hphantom{ }\textsf{cons}\hphantom{ }\symbol{58}\hphantom{ }\textsf{A}\hphantom{ }${\rightarrow}$\hphantom{ }\textsf{List}\hphantom{ }\textsf{A}\hphantom{ }${\rightarrow}$\hphantom{ }\textsf{List}\hphantom{ }\textsf{A}\symbol{46}\\
      \textbf{\textsf{Inductive}}\hphantom{ }\textsf{Tree}\hphantom{ }\symbol{58}\hphantom{ }\textbf{\textsf{Type}}\hphantom{ }\symbol{58}\symbol{61}\\
      ${\mid}$\hphantom{ }\textsf{node}\hphantom{ }\symbol{58}\hphantom{ }\textsf{List}\hphantom{ }\textsf{Tree}\hphantom{ }${\rightarrow}$\hphantom{ }\textsf{Tree}\symbol{46}
    \end{flushleft}
    }
  \end{displaymath}
  are indeed modelled in the algebraic presentation. 
  \par
  \subsection{Regular types}
  \par
  To be able to traverse terms of inductive type, the core \textsc{pts} constructions is extended with a \emph{recursive fixed point} on functions:
  \begin{displaymath}
    \parbox{1.\textwidth}{
    \begin{center}
      $\infer{\Gamma \vdash \mathrm{fix}~f~{x}_{1}{:}{A}_{1}~\mbox{\ldots }~{x}_{n-1}{:}{A}_{n-1}~{x}_{n}{:}{A}_{n}\Rightarrow u : \prod _{{x}_{1}{:}{A}_{1},\mbox{\ldots },{x}_{n-1}{:}{A}_{n-1},{x}_{n}{:}{A}_{n}}B}{
      \begin{array}{c}
        \eqmakebox[tag1]{$\Gamma \vdash \prod _{{x}_{1}{:}{A}_{1},\mbox{\ldots },{x}_{n-1}{:}{A}_{n-1},{x}_{n}{:}{A}_{n}}B : s$${\qquad}$\hspace{\stretch{1}}${\mathrm{guarded}}\,{f}\,{{x}_{1}}\,{\mbox{\ldots }}\,{{x}_{n-1}}\,{{x}_{n}}\,{u}$}\\
        \eqmakebox[tag1]{$\Gamma , f{:}\prod _{{x}_{1}{:}{A}_{1},\mbox{\ldots },{x}_{n-1}{:}{A}_{n-1},{x}_{n}{:}{A}_{n}}B, {x}_{1}{:}{A}_{1}, \mbox{\ldots }, {x}_{n-1}{:}{A}_{n-1}, {x}_{n}{:}{A}_{n}\vdash u : B$}\\
      \end{array}
      }$
    \end{center}
    }
  \end{displaymath}
  Recursive fixed points are unfolded when fully applied
  \begin{displaymath}
    \parbox{1.\textwidth}{
    \begin{center}
      ${\left( \mathrm{fix}~f~{x}_{1}{:}{A}_{1}~\mbox{\ldots }~{x}_{n}{:}{A}_{n}\Rightarrow u\right) }\,{{v}_{1}}\,{\mbox{\ldots }}\,{{v}_{n}}\leadsto {u}{\left[ \,{x}_{i}\,{\setminus }\,{v}_{i}\,\right] }$
    \end{center}
    }
  \end{displaymath}
  To ensure strong normalisation, this reduction rule is limited, and a guard condition is imposed on the recursive calls to $f$. It is not, however, the object of this article to discuss these restriction or the guard condition. Briefly, Coq currently relies on a single \emph{structural} argument in the block ${x}_{1}$, {\ldots}, ${x}_{n}$: fixed points are not unfolded until their structural argument starts with a constructor, and the guard condition ensures that each recursive call is performed on a subterm of said structural argument, for a relaxed notion of subterm. Other possibilities exist: Agda2~\cite{Norell2007} uses any number of arguments as structural, and tries to find a lexicographic ordering. Yet another possibilities is to use sized types~\cite{Abel2006}. We shall simply assume that an adequate guard condition is given.
  \par
  We now extend the grammar of type constructors. The presentation of this article is largely inspired by the synthetic definition of \emph{strictly positive families} by Morris \& al~\cite{Morris2007a,Morris2009}, but is adapted to \emph{intensional type theory}. The presentation of~\cite{Morris2007a,Morris2009} is designed for generic programming inside a type theory, they give codes for strictly positive families which are then decoded into an actual type of the ambient theory. No elimination principle needs to be given for the strictly positive families, as they are implicit in their decoding. Here, we are defining the syntax of inductive type families, including their elimination rules.
  \par
  \begin{figure}[tp]
    \centering{}\makebox[1.\textwidth]{\parbox{1.\textwidth}{
    \begin{flushleft}
      \textbf{Sum type}
    \end{flushleft}
    }}\\[0.5mm]
    \makebox[1.\textwidth]{\parbox{1.\textwidth}{
    \begin{center}
      $\infer{\Gamma \vdash A+B : \mathrm{Type}}{\Gamma \vdash A : \mathrm{Type} & \Gamma \vdash B : \mathrm{Type}}$
    \end{center}
    }}\\[0.5mm]
    \makebox[1.\textwidth]{\parbox{0.5\textwidth}{
    \begin{center}
      $\infer{\Gamma \vdash {\mathrm{inl}}\,{t} : A+B}{\Gamma \vdash A+B : s & \Gamma \vdash t : A}$
    \end{center}
    }\parbox{0.5\textwidth}{
    \begin{center}
      $\infer{\Gamma \vdash {\mathrm{inr}}\,{u} : A+B}{\Gamma \vdash A+B : s & \Gamma \vdash u : B}$
    \end{center}
    }}\\[0.5mm]
    \makebox[1.\textwidth]{\parbox{1.\textwidth}{
    \begin{center}
      $\infer{\Gamma \vdash 
      \begin{array}{@{}l}
        \mathrm{match}~u~\mathrm{as}~x~\mathrm{return}~P~\mathrm{with}\\
        \begin{array}{|l@{~\Rightarrow ~}l}
          {\mathrm{inl}}\,{y} & v\\
          {\mathrm{inr}}\,{z} & w\\
        \end{array}
        \\
      \end{array}
       : {P}{\left[ \,x\,{\setminus }\,u\,\right] }}{\Gamma \vdash u : A+B & \Gamma , x{:}A+B\vdash P : s & \Gamma , y{:}A\vdash v : {P}{\left[ \,x\,{\setminus }\,{\mathrm{inl}}\,{y}\,\right] } & \Gamma , z{:}A\vdash w : {P}{\left[ \,x\,{\setminus }\,{\mathrm{inr}}\,{z}\,\right] }}$
    \end{center}
    }}\\[0.5mm]
    \makebox[1.\textwidth]{\parbox{1.\textwidth}{
    \begin{center}
      $
      \begin{array}{@{}l}
        \mathrm{match}~{\mathrm{inl}}\,{u}~\mathrm{as}~x~\mathrm{return}~P~\mathrm{with}\\
        \begin{array}{|l@{~\Rightarrow ~}l}
          {\mathrm{inl}}\,{y} & v\\
          {\mathrm{inr}}\,{z} & w\\
        \end{array}
        \\
      \end{array}
      \leadsto {v}{\left[ \,y\,{\setminus }\,u\,\right] }$
    \end{center}
    }}\\[0.5mm]
    \makebox[1.\textwidth]{\parbox{1.\textwidth}{
    \begin{center}
      $
      \begin{array}{@{}l}
        \mathrm{match}~{\mathrm{inr}}\,{u}~\mathrm{as}~x~\mathrm{return}~P~\mathrm{with}\\
        \begin{array}{|l@{~\Rightarrow ~}l}
          {\mathrm{inl}}\,{y} & v\\
          {\mathrm{inr}}\,{z} & w\\
        \end{array}
        \\
      \end{array}
      \leadsto {w}{\left[ \,z\,{\setminus }\,u\,\right] }$
    \end{center}
    }}\\[0.5mm]
    \makebox[1.\textwidth]{\parbox{1.\textwidth}{
    \begin{flushleft}
      \textbf{Unit type}
    \end{flushleft}
    }}\\[0.5mm]
    \makebox[1.\textwidth]{\parbox{0.5\textwidth}{
    \begin{center}
      $\infer{\Gamma \vdash 1 : \mathrm{Type}}{}$
    \end{center}
    }\parbox{0.5\textwidth}{
    \begin{center}
      $\infer{\Gamma \vdash () : 1}{}$
    \end{center}
    }}\\[0.5mm]
    \makebox[1.\textwidth]{\parbox{1.\textwidth}{
    \begin{center}
      $\infer{\Gamma \vdash 
      \begin{array}{@{}l}
        \mathrm{match}~u~\mathrm{as}~x~\mathrm{return}~P~\mathrm{with}\\
        \begin{array}{|l@{~\Rightarrow ~}l}
          () & v\\
        \end{array}
        \\
      \end{array}
       : {P}{\left[ \,x\,{\setminus }\,u\,\right] }}{\Gamma \vdash u : 1 & \Gamma , x{:}1\vdash P : s & \Gamma \vdash v : {P}{\left[ \,x\,{\setminus }\,()\,\right] }}$
    \end{center}
    }}\\[0.5mm]
    \makebox[1.\textwidth]{\parbox{1.\textwidth}{
    \begin{center}
      $
      \begin{array}{@{}l}
        \mathrm{match}~()~\mathrm{as}~x~\mathrm{return}~P~\mathrm{with}\\
        \begin{array}{|l@{~\Rightarrow ~}l}
          () & v\\
        \end{array}
        \\
      \end{array}
      \leadsto v$
    \end{center}
    }}\\[0.5mm]
    \makebox[1.\textwidth]{\parbox{1.\textwidth}{
    \begin{flushleft}
      \textbf{Empty type}
    \end{flushleft}
    }}\\[0.5mm]
    \makebox[1.\textwidth]{\parbox{0.5\textwidth}{
    \begin{center}
      $\infer{\Gamma \vdash 0 : \mathrm{Type}}{}$
    \end{center}
    }\parbox{0.5\textwidth}{
    \begin{center}
      $\infer{\Gamma \vdash \mathrm{match}~u~\mathrm{return}~A~\mathrm{with}~\cdot  : A}{\Gamma \vdash u : 0 & \Gamma \vdash A : \mathrm{Type}}$
    \end{center}
    }}\\[0.5mm]
    \caption{Regular type constructors}\label{latex_lib_label_3}
  \end{figure}
  \par
  The regular type constructors, whose typing rules are given in Figure~\ref{latex_lib_label_3}, are the empty type $0$, the unit type $1$, and the sum of two types. The elimination rules are given in the form of \emph{dependent pattern-matching} with a syntax made to remind of that of Coq. We shall often omit the typing predicate when it is clear from the context, especially when it does not depend on the branch. With this material we can define a first example type, namely the booleans:
  \begin{displaymath}
    \left\{ 
    \begin{aligned}
      \mathbb{B}&=1+1\\
      \mathrm{true}&={\mathrm{inl}}\,{()}\\
      \mathrm{false}&={\mathrm{inr}}\,{()}
    \end{aligned}
    \right. 
  \end{displaymath}
  \par
  \subsection{Inductive type families}
  \par
  Inductive families differ from regular inductive types in that they are parametrised by \emph{indices}, that is they are functions $F : A\rightarrow \mathrm{Type}$ for some $A$. An inductive family of the form $\lambda {x}^{A}.\,R$, is said to be \emph{uniformly parametrised} by $A$. In general, inductive families are not uniformly parametrised: the value of the index is allowed to vary in recursive calls, and constructors may build values of ${F}\,{x}$ for certain $x$ only. Remember, for instance, the \textsf{EvenOdd} family:
  \begin{displaymath}
    \parbox{0.8\textwidth}{
    \begin{flushleft}
      \textbf{\textsf{Inductive}}\hphantom{ }\textsf{EvenOdd}\hphantom{ }\symbol{58}\hphantom{ }\textsf{bool}\hphantom{ }${\rightarrow}$\hphantom{ }\textbf{\textsf{Type}}\hphantom{ }\symbol{58}\symbol{61}\\
      ${\mid}$\hphantom{ }\textsf{eo}\hphantom{ }\symbol{58}\hphantom{ }\textsf{EvenOdd}\hphantom{ }\textsf{true}\\
      ${\mid}$\hphantom{ }\textsf{es}\hphantom{ }\symbol{58}\hphantom{ }\textsf{EvenOdd}\hphantom{ }\textsf{false}\hphantom{ }${\rightarrow}$\hphantom{ }\textsf{EvenOdd}\hphantom{ }\textsf{true}\\
      ${\mid}$\hphantom{ }\textsf{os}\hphantom{ }\symbol{58}\hphantom{ }\textsf{EvenOdd}\hphantom{ }\textsf{true}\hphantom{ }${\rightarrow}$\hphantom{ }\textsf{EvenOdd}\hphantom{ }\textsf{false}\symbol{46}
    \end{flushleft}
    }
  \end{displaymath}
  \par
  The inductive family constructors, presented in Figure~\ref{latex_lib_label_2}, warrant individual discussion. First, notice that as a simplifying hypothesis, inductive families have exactly one index. This is, of course, not a limitation in expressive power as multiple indices can be encoded as a dependent sum, and the unit type allows us to write families without an index.
  \par
  \begin{figure}[tp]
    \centering{}\makebox[1.\textwidth]{\parbox{1.\textwidth}{
    \begin{flushleft}
      \textbf{Inductive fixed point}
    \end{flushleft}
    }}\\[0.5mm]
    \makebox[1.\textwidth]{\parbox{1.\textwidth}{
    \begin{center}
      $\infer{\Gamma \vdash \mu {X}^{A\rightarrow \mathrm{Type}}.\,F : A\rightarrow \mathrm{Type}}{\Gamma \vdash A : s & \Gamma , X{:}A\rightarrow \mathrm{Type}\vdash F : A\rightarrow \mathrm{Type} & {{\mbox{\textsf{sp}}}_{X}}\,{F}}$
    \end{center}
    }}\\[0.5mm]
    \makebox[1.\textwidth]{\parbox{1.\textwidth}{
    \begin{center}
      $\mu {X}^{A\rightarrow s}.\,F\equiv {F}{\left[ \,X\,{\setminus }\,\mu {X}^{A\rightarrow s}.\,F\,\right] }$
    \end{center}
    }}\\[0.5mm]
    \makebox[1.\textwidth]{\parbox{1.\textwidth}{
    \begin{flushleft}
      \textbf{Proper indices}
    \end{flushleft}
    }}\\[0.5mm]
    \makebox[1.\textwidth]{\parbox{1.\textwidth}{
    \begin{center}
      $\infer{\Gamma \vdash {\sum }_{x{:}A}^{f}T : B\rightarrow \mathrm{Type}}{\Gamma \vdash A : \mathrm{Type} & \Gamma \vdash B : \mathrm{Type} & \Gamma , x{:}A\vdash T : \mathrm{Type} & \Gamma , x{:}A\vdash f : B}$
    \end{center}
    }}\\[0.5mm]
    \makebox[1.\textwidth]{\parbox{1.\textwidth}{
    \begin{center}
      $\infer{\Gamma \vdash {\left( u, v\right) }_{x{:}A.T}^{f} : {\left( {\sum }_{x{:}A}^{f}T\right) }\,{\left( {f}{\left[ \,x\,{\setminus }\,u\,\right] }\right) }}{\Gamma \vdash {\sum }_{x{:}A}^{f}T : B\rightarrow s & \Gamma \vdash u : A & \Gamma \vdash v : {T}{\left[ \,x\,{\setminus }\,u\,\right] }}$
    \end{center}
    }}\\[0.5mm]
    \makebox[1.\textwidth]{\parbox{1.\textwidth}{
    \begin{center}
      $\infer{\Gamma \vdash 
      \begin{array}{@{}l}
        \mathrm{match}~u~\mathrm{as}~z~\mathrm{in}~y~\mathrm{return}~P~\mathrm{with}\\
        \begin{array}{|l@{~\Rightarrow ~}l}
          {\left( i, j\right) }_{x{:}A.T}^{f} & v\\
        \end{array}
        \\
      \end{array}
       : {P}{\left[ \,y\,{\setminus }\,b\,,\,z\,{\setminus }\,u\,\right] }}{
      \begin{array}{c}
        \eqmakebox[tag2]{$\Gamma \vdash u : {\left( {\sum }_{x{:}A}^{f}T\right) }\,{b}$${\qquad}$\hspace{\stretch{1}}$\Gamma , y{:}B, z{:}{\left( {\sum }_{x{:}A}^{f}T\right) }\,{y}\vdash P : s$}\\
        \eqmakebox[tag2]{$\Gamma , i{:}A, j{:}{T}\,{i}\vdash v : {P}{\left[ \,y\,{\setminus }\,{f}\,{i}\,,\,z\,{\setminus }\,{\left( i, j\right) }_{x{:}A.T}^{f}\,\right] }$}\\
      \end{array}
      }$
    \end{center}
    }}\\[0.5mm]
    \makebox[1.\textwidth]{\parbox{1.\textwidth}{
    \begin{center}
      $
      \begin{array}{@{}l}
        \mathrm{match}~{\left( u, v\right) }_{x{:}A.T}^{f}~\mathrm{as}~z~\mathrm{in}~y~\mathrm{return}~P~\mathrm{with}\\
        \begin{array}{|l@{~\Rightarrow ~}l}
          {\left( i, j\right) }_{x{:}A.T}^{f} & w\\
        \end{array}
        \\
      \end{array}
      \leadsto {w}{\left[ \,i\,{\setminus }\,u\,,\,j\,{\setminus }\,v\,\right] }$
    \end{center}
    }}\\[0.5mm]
    \caption{Inductive type families}\label{latex_lib_label_2}
  \end{figure}
  \par
  \begin{figure}[tp]
    \centering{}\makebox[1.\textwidth]{\parbox{0.333333333333\textwidth}{
    \begin{center}
      $\infer{{{\mbox{\textsf{sp}}}_{X}}\,{y}}{}$
    \end{center}
    }\parbox{0.333333333333\textwidth}{
    \begin{center}
      $\infer{{{\mbox{\textsf{sp}}}_{X}}\,{0}}{}$
    \end{center}
    }\parbox{0.333333333333\textwidth}{
    \begin{center}
      $\infer{{{\mbox{\textsf{sp}}}_{X}}\,{1}}{}$
    \end{center}
    }}\\
    \makebox[1.\textwidth]{\parbox{0.333333333333\textwidth}{
    \begin{center}
      $\infer{{{\mbox{\textsf{sp}}}_{X}}\,{\left( \prod _{x{:}A}B\right) }}{X\mbox{\textrm{ is fresh in }}A & {{\mbox{\textsf{sp}}}_{X}}\,{B}}$
    \end{center}
    }\parbox{0.333333333333\textwidth}{
    \begin{center}
      $\infer{{{\mbox{\textsf{sp}}}_{X}}\,{\left( A+B\right) }}{{{\mbox{\textsf{sp}}}_{X}}\,{A} & {{\mbox{\textsf{sp}}}_{X}}\,{B}}$
    \end{center}
    }\parbox{0.333333333333\textwidth}{
    \begin{center}
      $\infer{{{\mbox{\textsf{sp}}}_{X}}\,{\left( \mu {Y}^{A}.\,F\right) }}{X\mbox{\textrm{ is fresh in }}A & {{\mbox{\textsf{sp}}}_{X}}\,{F}}$
    \end{center}
    }}\\
    \makebox[1.\textwidth]{\parbox{0.333333333333\textwidth}{
    \begin{center}
      $\infer{{{\mbox{\textsf{sp}}}_{X}}\,{\left( \lambda {x}^{A}.\,T\right) }}{X\mbox{\textrm{ is fresh in }}A & {{\mbox{\textsf{sp}}}_{X}}\,{T}}$
    \end{center}
    }\parbox{0.333333333333\textwidth}{
    \begin{center}
      $\infer{{{\mbox{\textsf{sp}}}_{X}}\,{\left( {U}\,{t}\right) }}{{{\mbox{\textsf{sp}}}_{X}}\,{U} & X\mbox{\textrm{ is fresh in }}t}$
    \end{center}
    }\parbox{0.333333333333\textwidth}{
    \begin{center}
      $\infer{{{\mbox{\textsf{sp}}}_{X}}\,{A}}{{{\mbox{\textsf{sp}}}_{X}}\,{B} & B\equiv A}$
    \end{center}
    }}\\
    \makebox[1.\textwidth]{\parbox{1.\textwidth}{
    \begin{center}
      $\infer{{{\mbox{\textsf{sp}}}_{X}}\,{\left( {\sum }_{x{:}A}^{f}T\right) }}{X\mbox{\textrm{ is fresh in }}f & {{\mbox{\textsf{sp}}}_{X}}\,{A} & {{\mbox{\textsf{sp}}}_{X}}\,{T}}$
    \end{center}
    }}\\
    \caption{Strict positivity condition}\label{latex_lib_label_4}
  \end{figure}
  \par
  The construction $\mu {X}^{A\rightarrow \mathrm{Type}}.\,F$ constructs the \emph{inductive fixed point} of $F$. It acts on type families, because indices vary through recursive calls to $X$. To be able to form an inductive fixed point, occurrences of $X$ must be strictly positive in $F$, rules for strict positivity are given in Figure~\ref{latex_lib_label_4}. The rules of Figure~\ref{latex_lib_label_4} are a simple set which suits the needs of this article, however in practice, we may want to consider strict positivity rules involving elimination rules and a finer treatment of application. Strict positivity ensures that no non-terminating term can be written without recursive fixed points, so that the guard condition suffices to enforce termination. Paradoxes which can be derived from non-positive or non-strictly positive inductive fixed points can be found in~\cite[Chapter 4, Section 4.2]{Paulin-Mohring1996}\cite[Chapter 3]{Morris2007}\cite[Chapter 8]{barras2013}.
  To avoid clutter, we give a presentation where inductive fixed points can be freely rolled and unrolled thanks to the conversion. An alternative can be to give an explicit term constructor for fixed points, see Section~\ref{latex_lib_label_9}.
  \par
  We will also use an inductive fixed point on nullary families, defined as:
  \begin{displaymath}
    \begin{aligned}
      \mu {X}^{\mathrm{Type}}.\,F&={\left( \mu {Y}^{1\rightarrow \mathrm{Type}}.\,{F}{\left[ \,X\,{\setminus }\,{Y}\,{()}\,\right] }\right) }\,{()}
    \end{aligned}
  \end{displaymath}
  from which we have that $\mu {X}^{\mathrm{Type}}.\,F$ can be freely rolled from or unrolled to
  ${F}{\left[ \,X\,{\setminus }\,\mu {X}^{\mathrm{Type}}.\,F\,\right] }$.
  \par
  With inductive fixed points, we can, for instance, define the accessibility predicate. In Coq:
  \begin{displaymath}
    \parbox{0.8\textwidth}{
    \begin{flushleft}
      \textbf{\textsf{Inductive}}\hphantom{ }\textsf{Acc}\hphantom{ }\symbol{40}\textsf{A}\symbol{58}\textbf{\textsf{Type}}\symbol{41}\hphantom{ }\symbol{40}\textsf{R}\symbol{58}\textsf{A}${\rightarrow}$\textsf{A}${\rightarrow}$\textbf{\textsf{Type}}\symbol{41}\hphantom{ }\symbol{40}\textsf{x}\symbol{58}\textsf{A}\symbol{41}\hphantom{ }\symbol{58}\hphantom{ }\textbf{\textsf{Type}}\hphantom{ }\symbol{58}\symbol{61}\\
      ${\mid}$\hphantom{ }\textsf{acc\_intro}\hphantom{ }\symbol{58}\hphantom{ }\symbol{40}\textbf{\textsf{forall}}\hphantom{ }\textsf{y}\symbol{58}\textsf{A}\symbol{44}\hphantom{ }\textsf{R}\hphantom{ }\textsf{y}\hphantom{ }\textsf{x}\hphantom{ }${\rightarrow}$\hphantom{ }\textsf{Acc}\hphantom{ }\textsf{A}\hphantom{ }\textsf{R}\hphantom{ }\textsf{y}\symbol{41}\hphantom{ }${\rightarrow}$\hphantom{ }\textsf{Acc}\hphantom{ }\textsf{A}\hphantom{ }\textsf{R}\hphantom{ }\textsf{x}\symbol{46}
    \end{flushleft}
    }
  \end{displaymath}
  This type represents the generic form of termination proofs: any terminating recursive fixed point can be made structural over a proof of accessibility. In the algebraic presentation, it is defined as:
  \begin{displaymath}
    \left\{ 
    \begin{aligned}
      \mathrm{Acc}&=\lambda {A}^{\mathrm{Type}}\,{R}^{A\rightarrow A\rightarrow \mathrm{Type}}.\,\mu {\mathrm{Acc}}^{A\rightarrow \mathrm{Type}}.\,\lambda {x}^{A}.\,\prod _{y{:}A}{R}\,{y}\,{x}\rightarrow {\mathrm{Acc}}\,{y}\\
      {\mathrm{acc}}_{\mathrm{intro}}&=\lambda {A}^{\mathrm{Type}}\,{R}^{A\rightarrow A\rightarrow \mathrm{Type}}\,{x}^{A}\,{f}^{\prod _{y{:}A}{R}\,{y}\,{x}\rightarrow {\mathrm{Acc}}\,{A}\,{R}\,{y}}.\,f
    \end{aligned}
    \right. 
  \end{displaymath}
  Because inductive fixed points are treated transparently, the constructor is rather trivial. However, notice how, in the definition of $\mathrm{Acc}$, the parameter $x$ is treated differently from $A$ and $R$. The reason is that $A$ and $R$ are \emph{uniform parameters}, in that they do not vary through recursive calls, whereas $x$ does: it is a \emph{non-uniform parameter}. The parameter $x$ is, hence, represented as an index. However, such an index is not sufficient to encode types like \textsf{EvenOdd}.
  \par
  Representing proper indices requires a new type construction, which we write ${\sum }_{x{:}A}^{f}T$. This construction comes from~\cite{Morris2007a,Morris2009}, where it is inspired by a categorical point of view: in a sufficiently extensional setting, ${\sum }_{x{:}A}^{f}T$ is the right adjoint to a pullback functor. The similarity with the usual notation of dependent sum is not fortuitous, indeed we can define dependent sum as a special case of proper indexing:
  \begin{displaymath}
    \left\{ 
    \begin{aligned}
      \sum _{x{:}A}B&={\left( {\sum }_{x{:}A}^{()}B\right) }\,{()}\\
      {\left( u, v\right) }_{{\mbox{\scriptsize{$x${:}$A$.$B$}}}}&={\left( u, v\right) }_{x{:}A.B}^{()}
    \end{aligned}
    \right. 
  \end{displaymath}
  We also write $A\times B$ and $\left( u,v\right) $ for $\sum _{x{:}A}B$ and ${\left( u, v\right) }_{{\mbox{\scriptsize{$x${:}$A$.$B$}}}}$, respectively, when $x$ is not free in $B$.
  \par
  In the case of dependent sums, the index is trivial. When it is not, however, the pattern matching return clause $P$ is allowed to depend on the value of the index. This is the purpose of Coq's \textsf{in}-pattern. With the algebraic presentation, the \textsf{in}-pattern has the pleasant property of being confined to the proper indexing construction, hopefully making its meaning more explicit. The syntax differs a little from that of Coq, however: Coq renders the \textsf{in} clause as a pattern with the type name at the head:
  \begin{displaymath}
    \parbox{0.8\textwidth}{
    \begin{flushleft}
      \textbf{\textsf{match}}\hphantom{ }\textsf{n}\hphantom{ }\textsf{as}\hphantom{ }\textsf{n}\symbol{39}\hphantom{ }\textsf{in}\hphantom{ }\textsf{EvenOdd}\hphantom{ }\textsf{b}\hphantom{ }\textbf{\textsf{return}}\hphantom{ }\textsf{P}\hphantom{ }\textsf{n}\symbol{39}\hphantom{ }\textsf{b}\hphantom{ }\textbf{\textsf{with}}\\
      {\ldots}\\
      \textbf{\textsf{end}}\symbol{46}
    \end{flushleft}
    }
  \end{displaymath}
  In the algebraic presentation, types not having a name, the \textsf{in} clause simply consists of a name for the index.
  \par
  The prototype of proper indexing is the identity type, which we name $\mathrm{Eq}$. In Coq:
  \begin{displaymath}
    \parbox{0.8\textwidth}{
    \begin{flushleft}
      \textbf{\textsf{Inductive}}\hphantom{ }\textsf{Eq}\hphantom{ }\symbol{40}\textsf{A}\symbol{58}\textbf{\textsf{Type}}\symbol{41}\hphantom{ }\symbol{40}\textsf{x}\symbol{58}\textsf{A}\symbol{41}\hphantom{ }\symbol{58}\hphantom{ }\textsf{A}\hphantom{ }${\rightarrow}$\hphantom{ }\textbf{\textsf{Type}}\hphantom{ }\symbol{58}\symbol{61}\\
      ${\mid}$\hphantom{ }\textsf{eq\_refl}\hphantom{ }\symbol{58}\hphantom{ }\textsf{Eq}\hphantom{ }\textsf{A}\hphantom{ }\textsf{x}\hphantom{ }\textsf{x}\\
    \end{flushleft}
    }
  \end{displaymath}
  in the algebraic presentation:
  \begin{displaymath}
    \left\{ 
    \begin{aligned}
      \mathrm{Eq}&=\lambda {A}^{\mathrm{Type}}\,{x}^{A}.\,{\sum }_{\_{:}1}^{x}1\\
      {\mathrm{eq}}_{\mathrm{refl}}&=\lambda {A}^{\mathrm{Type}}\,{x}^{A}.\,{\left( (), ()\right) }_{\_{:}1.1}^{x}
    \end{aligned}
    \right. 
  \end{displaymath}
  \par
  In fact, dependent sums and identity types are sufficient to define proper indexing. Indeed ${\sum }_{x{:}A}^{f}T$ can be redefined as:
  \begin{displaymath}
    \left\{ 
    \begin{aligned}
      {\sum }_{x{:}A}^{f}T&=\lambda {y}^{B}.\,\sum _{x{:}A}\left( {\mathrm{Eq}}\,{B}\,{y}\,{f}\right) \times T\\
      {\left( u, v\right) }_{x{:}A.T}^{f}&={\left( u, \left( {{\mathrm{eq}}_{\mathrm{refl}}}\,{B}\,{{f}{\left[ \,x\,{\setminus }\,u\,\right] }},v\right) \right) }_{{\mbox{\scriptsize{$x${:}$A$.$\left( {\mathrm{Eq}}\,{B}\,{\left( {f}{\left[ \,x\,{\setminus }\,u\,\right] }\right) }\,{f}\right) $${\times}$$T$}}}}
    \end{aligned}
    \right. 
  \end{displaymath}
  It is closer to the spirit of Coq, but in no way essential, to take a proper indexing construction rather than equality as primitive. In Morris \& al~\cite{Morris2007a,Morris2009}, the dependent sum and equality of the ambient type theory is used to define ${\sum }_{x{:}A}^{f}T$ which is then taken as primitive.
  \par
  An other choice lies in the use of $A+B$ as primitive. It is the only type construction which allows to define a type with distinct elements. However, a common alternative is to take $\mathbb{B}$ as primitive, in which case we can define $A+B$ as:
  \begin{displaymath}
    \left\{ 
    \begin{aligned}
      A+B&=\sum _{b{:}\mathbb{B}}
      \begin{array}{@{}l}
        \mathrm{match}~b~\mathrm{with}\\
        \begin{array}{|l@{~\Rightarrow ~}l}
          \mathrm{true} & A\\
          \mathrm{false} & B\\
        \end{array}
        \\
      \end{array}
      \\
      \mathrm{inl}&=\lambda {x}^{A}.\,{\left( \mathrm{true}, x\right) }_{{\mbox{\scriptsize{$b${:}$\mathbb{B}$.$
      \begin{array}{@{}l}
        \mathrm{match}~b~\mathrm{with}\\
        \begin{array}{|l@{~\Rightarrow ~}l}
          \mathrm{true} & A\\
          \mathrm{false} & B\\
        \end{array}
        \\
      \end{array}
      $}}}}\\
      \mathrm{inr}&=\lambda {y}^{B}.\,{\left( \mathrm{false}, y\right) }_{{\mbox{\scriptsize{$b${:}$\mathbb{B}$.$
      \begin{array}{@{}l}
        \mathrm{match}~b~\mathrm{with}\\
        \begin{array}{|l@{~\Rightarrow ~}l}
          \mathrm{true} & A\\
          \mathrm{false} & B\\
        \end{array}
        \\
      \end{array}
      $}}}}
    \end{aligned}
    \right. 
  \end{displaymath}
  \par
  \subsection{Examples}
  \par
  The previous section concludes the description of the predicative fragment of the algebraic presentation. We can now give definitions of the remaining inductive families we have encountered in terms of the algebraic presentation. Starting with \textsf{EvenOdd}:
  \begin{displaymath}
    \parbox{0.8\textwidth}{
    \begin{flushleft}
      \textbf{\textsf{Inductive}}\hphantom{ }\textsf{EvenOdd}\hphantom{ }\symbol{58}\hphantom{ }\textsf{bool}\hphantom{ }${\rightarrow}$\hphantom{ }\textbf{\textsf{Type}}\hphantom{ }\symbol{58}\symbol{61}\\
      ${\mid}$\hphantom{ }\textsf{eo}\hphantom{ }\symbol{58}\hphantom{ }\textsf{EvenOdd}\hphantom{ }\textsf{true}\\
      ${\mid}$\hphantom{ }\textsf{es}\hphantom{ }\symbol{58}\hphantom{ }\textsf{EvenOdd}\hphantom{ }\textsf{false}\hphantom{ }${\rightarrow}$\hphantom{ }\textsf{EvenOdd}\hphantom{ }\textsf{true}\\
      ${\mid}$\hphantom{ }\textsf{os}\hphantom{ }\symbol{58}\hphantom{ }\textsf{EvenOdd}\hphantom{ }\textsf{true}\hphantom{ }${\rightarrow}$\hphantom{ }\textsf{EvenOdd}\hphantom{ }\textsf{false}\symbol{46}
    \end{flushleft}
    }
  \end{displaymath}
  translates, in the algebraic presentation, to:
  \begin{displaymath}
    \left\{ 
    \begin{aligned}
      \mathrm{EvenOdd}&=\mu {\mathrm{EvenOdd}}^{\mathbb{B}\rightarrow \mathrm{Type}}.\,\left( {\sum }_{\_{:}1}^{\mathrm{true}}1+{\mathrm{EvenOdd}}\,{\mathrm{false}}\right) +\left( {\sum }_{\_{:}1}^{\mathrm{false}}{\mathrm{EvenOdd}}\,{\mathrm{true}}\right) \\
      \mathrm{eo}&={\mathrm{inl}}\,{{\left( (), {\mathrm{inl}}\,{()}\right) }_{\_{:}1.1}^{\mathrm{true}}}\\
      \mathrm{es}&=\lambda {x}^{{\mathrm{EvenOdd}}\,{\mathrm{false}}}.\,{\mathrm{inl}}\,{{\left( (), {\mathrm{inr}}\,{x}\right) }_{\_{:}1.1}^{\mathrm{true}}}\\
      \mathrm{os}&=\lambda {x}^{{\mathrm{EvenOdd}}\,{\mathrm{true}}}.\,{\mathrm{inr}}\,{{\left( (), x\right) }_{\_{:}1.1}^{\mathrm{false}}}
    \end{aligned}
    \right. 
  \end{displaymath}
  Here is the definition of \textsf{List} in Coq:
  \begin{displaymath}
    \parbox{0.8\textwidth}{
    \begin{flushleft}
      \textbf{\textsf{Inductive}}\hphantom{ }\textsf{List}\hphantom{ }\symbol{40}\textsf{A}\symbol{58}\textbf{\textsf{Type}}\symbol{41}\hphantom{ }\symbol{58}\hphantom{ }\textbf{\textsf{Type}}\hphantom{ }\symbol{58}\symbol{61}\\
      ${\mid}$\hphantom{ }\textsf{nil}\hphantom{ }\symbol{58}\hphantom{ }\textsf{List}\hphantom{ }\textsf{A}\\
      ${\mid}$\hphantom{ }\textsf{cons}\hphantom{ }\symbol{58}\hphantom{ }\textsf{A}\hphantom{ }${\rightarrow}$\hphantom{ }\textsf{List}\hphantom{ }\textsf{A}\hphantom{ }${\rightarrow}$\hphantom{ }\textsf{List}\hphantom{ }\textsf{A}\symbol{46}
    \end{flushleft}
    }
  \end{displaymath}
  and in the algebraic presentation:
  \begin{displaymath}
    \left\{ 
    \begin{aligned}
      \mathrm{List}&=\lambda {A}^{\mathrm{Type}}.\,\mu {\mathrm{List}}^{\mathrm{Type}}.\,1+A\times \mathrm{List}\\
      \mathrm{nil}&=\lambda {A}^{\mathrm{Type}}.\,{\mathrm{inl}}\,{()}\\
      \mathrm{cons}&=\lambda {A}^{\mathrm{Type}}\,{x}^{A}\,{l}^{{\mathrm{List}}\,{A}}.\,{\mathrm{inr}}\,{\left( x,l\right) }
    \end{aligned}
    \right. 
  \end{displaymath}
  Finally the definition of \textsf{Tree}:
  \begin{displaymath}
    \parbox{0.8\textwidth}{
    \begin{flushleft}
      \textbf{\textsf{Inductive}}\hphantom{ }\textsf{Tree}\hphantom{ }\symbol{58}\hphantom{ }\textbf{\textsf{Type}}\hphantom{ }\symbol{58}\symbol{61}\\
      ${\mid}$\hphantom{ }\textsf{node}\hphantom{ }\symbol{58}\hphantom{ }\textsf{List}\hphantom{ }\textsf{Tree}\hphantom{ }${\rightarrow}$\hphantom{ }\textsf{Tree}\symbol{46}
    \end{flushleft}
    }
  \end{displaymath}
  translates to:
  \begin{displaymath}
    \left\{ 
    \begin{aligned}
      \mathrm{Tree}&=\mu {\mathrm{Tree}}^{\mathrm{Type}}.\,{\mathrm{List}}\,{\mathrm{Tree}}\\
      \mathrm{node}&=\lambda {f}^{{\mathrm{List}}\,{\mathrm{Tree}}}.\,f
    \end{aligned}
    \right. 
  \end{displaymath}
  Note that in the definition of $\mathrm{Tree}$, we must ${\beta}$-reduce ${\mathrm{List}}\,{\mathrm{Tree}}$ for the recursive definition to be strictly positive.
  \par
  A more complex example is given by the type of lists indexed by their length, often called \emph{vectors}:
  \begin{displaymath}
    \parbox{0.8\textwidth}{
    \begin{flushleft}
      \textbf{\textsf{Inductive}}\hphantom{ }\textsf{Nat}\hphantom{ }\symbol{58}\hphantom{ }\textbf{\textsf{Type}}\hphantom{ }\symbol{58}\symbol{61}\\
      ${\mid}$\hphantom{ }\textsf{o}\hphantom{ }\symbol{58}\hphantom{ }\textsf{Nat}\\
      ${\mid}$\hphantom{ }\textsf{s}\hphantom{ }\symbol{58}\hphantom{ }\textsf{Nat}\hphantom{ }${\rightarrow}$\hphantom{ }\textsf{Nat}\symbol{46}\\
      \textbf{\textsf{Inductive}}\hphantom{ }\textsf{Vector}\hphantom{ }\symbol{40}\textsf{A}\symbol{58}\textbf{\textsf{Type}}\symbol{41}\hphantom{ }\symbol{58}\hphantom{ }\textsf{Nat}\hphantom{ }${\rightarrow}$\hphantom{ }\textbf{\textsf{Type}}\hphantom{ }\symbol{58}\symbol{61}\\
      ${\mid}$\hphantom{ }\textsf{vnil}\hphantom{ }\symbol{58}\hphantom{ }\textsf{Vector}\hphantom{ }\textsf{A}\hphantom{ }\textsf{o}\\
      ${\mid}$\hphantom{ }\textsf{vcons}\hphantom{ }\symbol{58}\hphantom{ }\textbf{\textsf{forall}}\hphantom{ }\textsf{n}\symbol{44}\hphantom{ }\textsf{A}\hphantom{ }${\rightarrow}$\hphantom{ }\textsf{Vector}\hphantom{ }\textsf{A}\hphantom{ }\textsf{n}\hphantom{ }${\rightarrow}$\hphantom{ }\textsf{Vector}\hphantom{ }\textsf{A}\hphantom{ }\symbol{40}\textsf{s}\hphantom{ }\textsf{n}\symbol{41}\symbol{46}\\
    \end{flushleft}
    }
  \end{displaymath}
  It is encoded in the algebraic presentation as:
  \begin{displaymath}
    \left\{ 
    \begin{aligned}
      \mathrm{Nat}&=\mu {\mathrm{Nat}}^{\mathrm{Type}}.\,1+\mathrm{Nat}\\
      o&={\mathrm{inl}}\,{()}\\
      s&=\lambda {n}^{\mathrm{Nat}}.\,{\mathrm{inr}}\,{n}\\
      \mathrm{Vector}&=\lambda {A}^{\mathrm{Type}}.\,\mu {V}^{\mathrm{Nat}\rightarrow \mathrm{Type}}.\,\lambda {n}^{\mathrm{Nat}}.\,{\left( {\sum }_{\_{:}1}^{o}1\right) }\,{n}+{\left( {\sum }_{n'{:}\mathrm{Nat}}^{{s}\,{n'}}A\times {V}\,{n'}\right) }\,{n}\\
      \mathrm{vnil}&=\lambda {A}^{\mathrm{Type}}.\,{\mathrm{inl}}\,{{\left( (), ()\right) }_{\_{:}1.1}^{o}}\\
      \mathrm{vcons}&=\lambda {A}^{\mathrm{Type}}\,{n}^{\mathrm{Nat}}\,{a}^{A}\,{v}^{{\mathrm{Vector}}\,{n}\,{a}}.\,{\mathrm{inr}}\,{{\left( n, \left( a,v\right) \right) }_{n'{:}\mathrm{Nat}.A\times {V}\,{n'}}^{{s}\,{n'}}}
    \end{aligned}
    \right. 
  \end{displaymath}
  \par
  Contrary to proper indices, the types of non-uniform parameters are allowed to be in $\Box $, this allows the definition of types such as the binary lists~\cite{Okasaki1999}:
  \begin{displaymath}
    \parbox{0.8\textwidth}{
    \begin{flushleft}
      \textbf{\textsf{Inductive}}\hphantom{ }\textsf{BList}\hphantom{ }\symbol{40}\textsf{A}\symbol{58}\textbf{\textsf{Type}}\symbol{41}\hphantom{ }\symbol{58}\hphantom{ }\textbf{\textsf{Type}}\hphantom{ }\symbol{58}\symbol{61}\\
      ${\mid}$\hphantom{ }\textsf{one}\hphantom{ }\symbol{58}\hphantom{ }\textsf{A}\hphantom{ }${\rightarrow}$\hphantom{ }\textsf{BList}\hphantom{ }\textsf{A}\\
      ${\mid}$\hphantom{ }\textsf{twice}\hphantom{ }\symbol{58}\hphantom{ }\textsf{BList}\hphantom{ }\symbol{40}\textsf{A}$*$\textsf{A}\symbol{41}\hphantom{ }${\rightarrow}$\hphantom{ }\textsf{BList}\hphantom{ }\textsf{A}\\
      ${\mid}$\hphantom{ }\textsf{stwice}\hphantom{ }\symbol{58}\hphantom{ }\textsf{A}\hphantom{ }${\rightarrow}$\hphantom{ }\textsf{BList}\hphantom{ }\symbol{40}\textsf{A}$*$\textsf{A}\symbol{41}\hphantom{ }${\rightarrow}$\hphantom{ }\textsf{BList}\hphantom{ }\textsf{A}
    \end{flushleft}
    }
  \end{displaymath}
  which are rendered in the algebraic presentation as:
  \begin{displaymath}
    \left\{ 
    \begin{aligned}
      \mathrm{BList}&=\mu {\mathrm{BList}}^{\mathrm{Type}\rightarrow \mathrm{Type}}.\,\lambda {A}^{\mathrm{Type}}.\,A+\left( {\mathrm{BList}}\,{\left( A\times A\right) }+A\times \left( {\mathrm{BList}}\,{\left( A\times A\right) }\right) \right) \\
      \mathrm{one}&=\lambda {A}^{\mathrm{Type}}\,{x}^{A}.\,{\mathrm{inl}}\,{x}\\
      \mathrm{twice}&=\lambda {A}^{\mathrm{Type}}\,{l}^{{\mathrm{BList}}\,{\left( A\times A\right) }}.\,{\mathrm{inr}}\,{\left( {\mathrm{inl}}\,{l}\right) }\\
      \mathrm{stwice}&=\lambda {A}^{\mathrm{Type}}\,{x}^{A}\,{l}^{{\mathrm{BList}}\,{\left( A\times A\right) }}.\,{\mathrm{inr}}\,{\left( {\mathrm{inr}}\,{\left( a,l\right) }\right) }
    \end{aligned}
    \right. 
  \end{displaymath}
  \par
  In Coq, where there is a hierarchy of universe, types of proper indices can be in any sort. However, a proper index whose type is in ${\mathrm{Type}}_{i}$ constrains the final type to be in ${\mathrm{Type}}_{i+1}$ or higher. Uniform parameters, of any type, do not constrain the type they parametrise.
  \par
  Inductive types are consumed by recursive fixed points. Using the implicit unfolding of inductive fixed points, we can pattern-match over the top constructor directly. The Coq function
  \begin{displaymath}
    \parbox{0.8\textwidth}{
    \begin{flushleft}
      \textbf{\textsf{Fixpoint}}\hphantom{ }\textsf{add}\hphantom{ }\symbol{40}\textsf{x}\hphantom{ }\textsf{y}\symbol{58}\textsf{Nat}\symbol{41}\hphantom{ }\symbol{58}\hphantom{ }\textsf{Nat}\hphantom{ }\symbol{58}\symbol{61}\\
      \textbf{\textsf{match}}\hphantom{ }\textsf{y}\hphantom{ }\textbf{\textsf{with}}\\
      ${\mid}$\hphantom{ }\textsf{o}\hphantom{ }${\Rightarrow}$\hphantom{ }\textsf{x}\\
      ${\mid}$\hphantom{ }\textsf{s}\hphantom{ }\textsf{y}\symbol{39}\hphantom{ }${\Rightarrow}$\hphantom{ }\textsf{s}\hphantom{ }\symbol{40}\textsf{add}\hphantom{ }\textsf{x}\hphantom{ }\textsf{y}\symbol{39}\symbol{41}\\
      \textbf{\textsf{end}}
    \end{flushleft}
    }
  \end{displaymath}
  is rendered in the algebraic presentation as
  \begin{displaymath}
    \begin{aligned}
      \mathrm{add}&=\mathrm{fix}~\mathrm{add}~x{:}\mathrm{Nat}~y{:}\mathrm{Nat}\Rightarrow 
      \begin{array}{@{}l}
        \mathrm{match}~y~\mathrm{as}~\_~\mathrm{return}~\mathrm{Nat}~\mathrm{with}\\
        \begin{array}{|l@{~\Rightarrow ~}l}
          {\mathrm{inl}}\,{\_} & x\\
          {\mathrm{inr}}\,{y'} & {s}\,{\left( {\mathrm{add}}\,{x}\,{y'}\right) }\\
        \end{array}
        \\
      \end{array}
    \end{aligned}
  \end{displaymath}
  \par
  \subsection{Co-induction}\label{latex_lib_label_9}
  \par
  In addition to inductive fixed points, Coq also has support for co-inductive fixed points. Co-inductive fixed points are required to be strictly positive, like inductive fixed points. We choose in this section, a presentation of co-inductive data where fixed points are explicitly introduced with a constructor. Below we will use this explicit presentation to give a variation on Coq's co-inductive fixed points.
  \begin{displaymath}
    \parbox{1.\textwidth}{
    \begin{center}
      \makebox[1.\textwidth]{\parbox{1.\textwidth}{
      \begin{center}
        $\infer{\Gamma \vdash \nu {X}^{A\rightarrow \mathrm{Type}}.\,F : A\rightarrow \mathrm{Type}}{\Gamma \vdash A : s & \Gamma , X{:}A\rightarrow \mathrm{Type}\vdash F : A\rightarrow \mathrm{Type} & {{\mbox{\textsf{sp}}}_{X}}\,{F}}$
      \end{center}
      }}\\[0.5mm]
      \makebox[1.\textwidth]{\parbox{1.\textwidth}{
      \begin{center}
        $\infer{\Gamma \vdash {\mathrm{forced}}\,{u} : {\left( \nu {X}^{A\rightarrow s}.\,F\right) }\,{i}}{\Gamma \vdash \nu {X}^{A\rightarrow s}.\,F : A\rightarrow s & \Gamma \vdash i : A & \Gamma \vdash u : {{F}{\left[ \,X\,{\setminus }\,\nu {X}^{A\rightarrow s}.\,F\,\right] }}\,{i}}$
      \end{center}
      }}\\[0.5mm]
      \makebox[1.\textwidth]{\parbox{1.\textwidth}{
      \begin{center}
        $\infer{\Gamma \vdash 
        \begin{array}{@{}l}
          \mathrm{match}~u~\mathrm{as}~x~\mathrm{return}~P~\mathrm{with}\\
          \begin{array}{|l@{~\Rightarrow ~}l}
            {\mathrm{forced}}\,{y} & v\\
          \end{array}
          \\
        \end{array}
         : {P}{\left[ \,x\,{\setminus }\,u\,\right] }}{
        \begin{array}{c}
          \eqmakebox[tag3]{$\Gamma \vdash \nu {X}^{A\rightarrow s}.\,F : A\rightarrow s$${\qquad}$\hspace{\stretch{1}}$\Gamma \vdash i : A$${\qquad}$\hspace{\stretch{1}}$\Gamma \vdash u : {\left( \nu {X}^{A\rightarrow s}.\,F\right) }\,{i}$}\\
          \eqmakebox[tag3]{$\Gamma , x{:}{\left( \nu {X}^{A\rightarrow s}.\,F\right) }\,{i}\vdash P : s'$${\qquad}$\hspace{\stretch{1}}$\Gamma , y{:}{{F}{\left[ \,X\,{\setminus }\,\nu {X}^{A\rightarrow s}.\,F\,\right] }}\,{i}\vdash v : {P}{\left[ \,x\,{\setminus }\,{\mathrm{forced}}\,{y}\,\right] }$}\\
        \end{array}
        }$
      \end{center}
      }}\\[0.5mm]
      \makebox[1.\textwidth]{\parbox{1.\textwidth}{
      \begin{center}
        $
        \begin{array}{@{}l}
          \mathrm{match}~{\mathrm{forced}}\,{u}~\mathrm{as}~x~\mathrm{return}~P~\mathrm{with}\\
          \begin{array}{|l@{~\Rightarrow ~}l}
            {\mathrm{forced}}\,{y} & v\\
          \end{array}
          \\
        \end{array}
        \leadsto {v}{\left[ \,y\,{\setminus }\,u\,\right] }$
      \end{center}
      }}\\[0.5mm]
    \end{center}
    }
  \end{displaymath}
  \par
  Just like inductive data is \emph{destructed} by a recursive fixed point operation, co-inductive data is \emph{constructed} by a co-recursive fixed point operation, allowing co-inductive data to be infinite. The guard condition on co-recursive fixed points ensures that a finite number of unfolding will eventually produce a forced value.
  \begin{displaymath}
    \parbox{1.\textwidth}{
    \begin{center}
      $\infer{\Gamma \vdash \mathrm{cofix}~f~{x}_{1}{:}{A}_{1}~\mbox{\ldots }~{x}_{n}{:}{A}_{n}\Rightarrow u : \prod _{{x}_{1}{:}{A}_{1},\mbox{\ldots },{x}_{n}{:}{A}_{n}}{\left( \nu {X}^{A\rightarrow s}.\,F\right) }\,{i}}{
      \begin{array}{c}
        \eqmakebox[tag4]{$\Gamma \vdash \prod _{{x}_{1}{:}{A}_{1},\mbox{\ldots },{x}_{n}{:}{A}_{n}}{\left( \nu {X}^{A\rightarrow s}.\,F\right) }\,{i} : s$${\qquad}$\hspace{\stretch{1}}${\mathrm{coguarded}}\,{f}\,{{x}_{1}}\,{\mbox{\ldots }}\,{{x}_{n}}\,{u}$}\\
        \eqmakebox[tag4]{$\Gamma , f{:}\prod _{{x}_{1}{:}{A}_{1},\mbox{\ldots },{x}_{n}{:}{A}_{n}}{\left( \nu {X}^{A\rightarrow s}.\,F\right) }\,{i}, {x}_{1}{:}{A}_{1}, \mbox{\ldots }, {x}_{n}{:}{A}_{n}\vdash u : {\left( \nu {X}^{A\rightarrow s}.\,F\right) }\,{i}$}\\
      \end{array}
      }$
    \end{center}
    }
  \end{displaymath}
  Co-recursive fixed-point are meant to represent infinite data: they cannot be unfolded eagerly, lest they would fail to terminate. They are unfolded only when they appear at the head of a pattern-matching expression:
  \begin{displaymath}
    \parbox{1.\textwidth}{
    \begin{center}
      $
      \begin{array}{ll}
         & 
        \begin{array}{@{}l}
          \mathrm{match}~{\left( \mathrm{cofix}~f~{x}_{1}{:}{A}_{1}~\mbox{\ldots }~{x}_{n}{:}{A}_{n}\Rightarrow u\right) }\,{{v}_{1}}\,{\mbox{\ldots }}\,{{v}_{n}}~\mathrm{as}~x~\mathrm{return}~P~\mathrm{with}\\
          \begin{array}{|l@{~\Rightarrow ~}l}
            {\mathrm{forced}}\,{y} & v\\
          \end{array}
          \\
        \end{array}
        \\
        \leadsto  & 
        \begin{array}{@{}l}
          \mathrm{match}~{u}{\left[ \,f\,{\setminus }\,\left( \mathrm{cofix}~f~{x}_{1}{:}{A}_{1}~\mbox{\ldots }~{x}_{n}{:}{A}_{n}\Rightarrow u\right) \,,\,{x}_{i}\,{\setminus }\,{v}_{i}\,\right] }~\mathrm{as}~x~\mathrm{return}~P~\mathrm{with}\\
          \begin{array}{|l@{~\Rightarrow ~}l}
            {\mathrm{forced}}\,{y} & v\\
          \end{array}
          \\
        \end{array}
        \\
      \end{array}
      $
    \end{center}
    }
  \end{displaymath}
  \par
  The dependent elimination rule for co-inductive fixed points asserts, in essence, that every co-inductive data is of the form ${\mathrm{forced}}\,{u}$. Even though it would be fine for inductive fixed points -- this is why we could leave the unrolling to the conversion -- this does not reflect well the computational aspects of co-inductive data: suspended co-recursive fixed points are values, and won't be evaluated until the context demands it. The fact that the elimination for co-inductive data claims that all values are forced gives rise to undesirable behaviour.
  \par
  Take for instance the following simple co-inductive type, and data:
  \begin{displaymath}
    \left\{ 
    \begin{aligned}
      T&=\nu X.\,X\\
      i&=\mathrm{cofix}~i\Rightarrow {\mathrm{forced}}\,{i}
    \end{aligned}
    \right. 
  \end{displaymath}
  So that $i$ is effectively an infinite sequence of $\mathrm{forced}$. Using the elimination principle above, it is possible to give a \emph{closed} proof that ${\mathrm{Eq}}\,{T}\,{i}\,{\left( {\mathrm{forced}}\,{i}\right) }$:
  \begin{displaymath}
    \parbox{0.8\textwidth}{
    \begin{flushleft}
      $
      \begin{array}{@{}l}
        \mathrm{match}~i~\mathrm{as}~x~\mathrm{return}~{\mathrm{Eq}}\,{T}\,{x}\,{\left( {\mathrm{forced}}\,{\left( 
        \begin{array}{@{}l}
          \mathrm{match}~x~\mathrm{with}\\
          \begin{array}{|l@{~\Rightarrow ~}l}
            {\mathrm{forced}}\,{y} & y\\
          \end{array}
          \\
        \end{array}
        \right) }\right) }~\mathrm{with}\\
        \begin{array}{|l@{~\Rightarrow ~}l}
          {\mathrm{forced}}\,{y} & {{\mathrm{eq}}_{\mathrm{refl}}}\,{T}\,{y}\\
        \end{array}
        \\
      \end{array}
      $
    \end{flushleft}
    }
  \end{displaymath}
   However, $i$ and $\left( {\mathrm{forced}}\,{i}\right) $ are not convertible, yet, as every closed proof of equality does, this proof reduces to ${\mathrm{eq}}_{\mathrm{refl}}$, hence should relate convertible terms. The dependent elimination rule of co-inductive fixed points compromises the type safety of the logic.
  \par
  Coq uses the above dependent elimination rule for co-inductive fixed points. It was a deliberate decision made for practical purposes. Nonetheless, one may want to weaken it to avoid the incompatibility between equality and conversion. To do so, it suffices to erase the dependency of the return predicate over the matched term:
  \begin{displaymath}
    \parbox{1.\textwidth}{
    \begin{center}
      $\infer{\Gamma \vdash 
      \begin{array}{@{}l}
        \mathrm{match}~u~\mathrm{return}~P~\mathrm{with}\\
        \begin{array}{|l@{~\Rightarrow ~}l}
          {\mathrm{forced}}\,{y} & v\\
        \end{array}
        \\
      \end{array}
       : P}{\Gamma \vdash \nu {X}^{A\rightarrow s}.\,F : A\rightarrow s & \Gamma \vdash i : A & \Gamma \vdash u : {\left( \nu {X}^{A\rightarrow s}.\,F\right) }\,{i} & \Gamma \vdash P : s' & \Gamma \vdash v : P}$
    \end{center}
    }
  \end{displaymath}
  \section{Prop}
  \par
  With all the common baggage for predicative sorts set in place, we can add impredicative sorts to the algebraic presentation. The main such sort in Coq is the sort \textbf{\textsf{Prop}} of propositions. The design of \textbf{\textsf{Prop}} is guided by \emph{proof irrelevance}: even if it is not actually provable in Coq, different proofs of a proposition are thought of as being equal. This property is useful for program extraction: only the \emph{computationally relevant} parts of a program need to be executed to get the final result. In other words: propositions are considered as \emph{static} data. It is why, with disjunction and existential defined as:
  \begin{displaymath}
    \parbox{0.8\textwidth}{
    \begin{flushleft}
      \textbf{\textsf{Inductive}}\hphantom{ }\textsf{Or}\hphantom{ }\symbol{40}\textsf{A}\hphantom{ }\textsf{B}\symbol{58}\textbf{\textsf{Prop}}\symbol{41}\hphantom{ }\symbol{58}\hphantom{ }\textbf{\textsf{Prop}}\hphantom{ }\symbol{58}\symbol{61}\\
      ${\mid}$\hphantom{ }\textsf{or\_introl}\hphantom{ }\symbol{58}\hphantom{ }\textsf{A}\hphantom{ }${\rightarrow}$\hphantom{ }\textsf{A}\hphantom{ }${\lor}$\hphantom{ }\textsf{B}\\
      ${\mid}$\hphantom{ }\textsf{or\_intror}\hphantom{ }\symbol{58}\hphantom{ }\textsf{B}\hphantom{ }${\rightarrow}$\hphantom{ }\textsf{A}\hphantom{ }${\lor}$\hphantom{ }\textsf{B}\symbol{46}\\
      \textbf{\textsf{Inductive}}\hphantom{ }\textsf{Ex}\hphantom{ }\symbol{40}\textsf{A}\symbol{58}\textbf{\textsf{Type}}\symbol{41}\hphantom{ }\symbol{40}\textsf{P}\symbol{58}\textsf{A}${\rightarrow}$\textbf{\textsf{Prop}}\symbol{41}\hphantom{ }\symbol{58}\hphantom{ }\textbf{\textsf{Prop}}\hphantom{ }\symbol{58}\symbol{61}\\
      ${\mid}$\hphantom{ }\textsf{ex\_intro}\hphantom{ }\symbol{58}\hphantom{ }\textbf{\textsf{forall}}\hphantom{ }\textsf{x}\symbol{58}\textsf{A}\symbol{44}\hphantom{ }\textsf{P}\hphantom{ }\textsf{x}\hphantom{ }${\rightarrow}$\hphantom{ }\textsf{Ex}\hphantom{ }\textsf{A}\hphantom{ }\textsf{P}\symbol{46}
    \end{flushleft}
    }
  \end{displaymath}
  the following terms are refused by type-checking:
  \begin{displaymath}
    \parbox{0.8\textwidth}{
    \begin{flushleft}
      \textbf{\textsf{match}}\hphantom{ }\textsf{x}\hphantom{ }\textbf{\textsf{with}}\\
      ${\mid}$\hphantom{ }\textsf{or\_introl}\hphantom{ }$\_$\hphantom{ }${\Rightarrow}$\hphantom{ }\textsf{true}\\
      ${\mid}$\hphantom{ }\textsf{or\_intror}\hphantom{ }$\_$\hphantom{ }${\Rightarrow}$\hphantom{ }\textsf{false}\\
      \textbf{\textsf{end}}\symbol{46}
    \end{flushleft}
    }
  \end{displaymath}
  and
  \begin{displaymath}
    \parbox{0.8\textwidth}{
    \begin{flushleft}
      \textbf{\textsf{match}}\hphantom{ }\textsf{x}\hphantom{ }\textbf{\textsf{with}}\\
      ${\mid}$\hphantom{ }\textsf{ex\_intro}\hphantom{ }\textsf{x}\hphantom{ }$\_$\hphantom{ }${\Rightarrow}$\hphantom{ }\textsf{x}\\
      \textbf{\textsf{end}}\symbol{46}
    \end{flushleft}
    }
  \end{displaymath}
  \par
  On the other hand, it is not the case of every inductive type defined in \textbf{\textsf{Prop}}, that they cannot be eliminated into \textbf{\textsf{Type}}. Conjunction and falsity are two counter-examples:
  \begin{displaymath}
    \parbox{0.8\textwidth}{
    \begin{flushleft}
      \textbf{\textsf{Inductive}}\hphantom{ }\textsf{False}\hphantom{ }\symbol{58}\hphantom{ }\textbf{\textsf{Prop}}\hphantom{ }\symbol{58}\symbol{61}\hphantom{ }\symbol{46}\\
      \textbf{\textsf{Inductive}}\hphantom{ }\textsf{And}\hphantom{ }\symbol{40}\textsf{A}\hphantom{ }\textsf{B}\symbol{58}\textbf{\textsf{Prop}}\symbol{41}\hphantom{ }\symbol{58}\hphantom{ }\textbf{\textsf{Prop}}\hphantom{ }\symbol{58}\symbol{61}\\
      ${\mid}$\hphantom{ }\textsf{conj}\hphantom{ }\symbol{58}\hphantom{ }\textsf{A}\hphantom{ }${\rightarrow}$\hphantom{ }\textsf{B}\hphantom{ }${\rightarrow}$\hphantom{ }\textsf{A}\hphantom{ }${\land}$\hphantom{ }\textsf{B}\symbol{46}
    \end{flushleft}
    }
  \end{displaymath}
  Coq allows elimination over these two propositions into \textbf{\textsf{Type}}, and both following terms are well-typed:
  \begin{displaymath}
    \parbox{0.8\textwidth}{
    \begin{flushleft}
      \textbf{\textsf{match}}\hphantom{ }\textsf{x}\hphantom{ }\textbf{\textsf{return}}\hphantom{ }\textsf{Bool}\hphantom{ }\textbf{\textsf{with}}\hphantom{ }\textbf{\textsf{end}}\symbol{46}
    \end{flushleft}
    }
  \end{displaymath}
  and
  \begin{displaymath}
    \parbox{0.8\textwidth}{
    \begin{flushleft}
      \textbf{\textsf{match}}\hphantom{ }\textsf{x}\hphantom{ }\textbf{\textsf{with}}\\
      ${\mid}$\hphantom{ }\textsf{conj}\hphantom{ }$\_$\hphantom{ }$\_$\hphantom{ }${\Rightarrow}$\hphantom{ }\textsf{true}\\
      \textbf{\textsf{end}}\symbol{46}
    \end{flushleft}
    }
  \end{displaymath}
  \par
  The object of this section is to make syntactically explicit what happens when an inductive type of Coq is declared to be of sort \textbf{\textsf{Prop}}. The description elaborated in this section has strong similarities with the system of bracket-types proposed by Awodey \& Bauer~\cite{Awodey2004}. They describe the propositions as the subset of types with at most one element, and introduce a left adjoint, written as brackets, to the inclusion of propositions into types. We will reuse their notation, even though, in our intensional setting, \textsf{T}\symbol{58}\textbf{\textsf{Prop}} does not enforce that \textsf{T} has a most one element, and the bracketing operation does not properly form an adjunction with the inclusion from \textbf{\textsf{Prop}} to \textbf{\textsf{Type}}.
  \par
  \subsection{Impredicativity}
  \par
  Let us start by introducing the new sort $\mathrm{Prop}$ in the algebraic presentation:
  \begin{displaymath}
    \parbox{1.\textwidth}{
    \begin{center}
      $\infer{\Gamma \vdash \mathrm{Prop} : \mathrm{Type}}{}$
    \end{center}
    }
  \end{displaymath}
  As in~\cite{Awodey2004}, propositions form a subset of types. Coq has a subtyping rule (also known as \emph{cumulativity}) to make the inclusion transparent. We will, however, render it with a syntactic construct:
  \begin{displaymath}
    \parbox{1.\textwidth}{
    \begin{center}
      \makebox[1.\textwidth]{\parbox{0.5\textwidth}{
      \begin{center}
        $\infer{\Gamma \vdash \left\{ A\right\}  : \mathrm{Type}}{\Gamma \vdash A : \mathrm{Prop}}$
      \end{center}
      }\parbox{0.5\textwidth}{
      \begin{center}
        $\infer{\Gamma \vdash {\mathrm{prf}}\,{u} : \left\{ A\right\} }{\Gamma \vdash u : A}$
      \end{center}
      }}\\[0.5mm]
      \makebox[1.\textwidth]{\parbox{1.\textwidth}{
      \begin{center}
        $\infer{\Gamma \vdash 
        \begin{array}{@{}l}
          \mathrm{match}~u~\mathrm{as}~x~\mathrm{return}~P~\mathrm{with}\\
          \begin{array}{|l@{~\Rightarrow ~}l}
            {\mathrm{prf}}\,{y} & v\\
          \end{array}
          \\
        \end{array}
         : {P}{\left[ \,x\,{\setminus }\,u\,\right] }}{\Gamma \vdash u : \left\{ A\right\}  & \Gamma , x{:}\left\{ A\right\} \vdash P : s & \Gamma , y{:}A\vdash v : {P}{\left[ \,x\,{\setminus }\,{\mathrm{prf}}\,{y}\,\right] }}$
      \end{center}
      }}\\[0.5mm]
      \makebox[1.\textwidth]{\parbox{1.\textwidth}{
      \begin{center}
        $
        \begin{array}{@{}l}
          \mathrm{match}~{\mathrm{prf}}\,{u}~\mathrm{as}~x~\mathrm{return}~P~\mathrm{with}\\
          \begin{array}{|l@{~\Rightarrow ~}l}
            {\mathrm{prf}}\,{y} & v\\
          \end{array}
          \\
        \end{array}
        \leadsto {v}{\left[ \,y\,{\setminus }\,u\,\right] }$
      \end{center}
      }}\\[0.5mm]
    \end{center}
    }
  \end{displaymath}
  This definition simply makes $\left\{ A\right\} $ a synonym of $A$, except of sort $\mathrm{Type}$. It is strictly positive in $A$:
  \begin{displaymath}
    \parbox{1.\textwidth}{
    \begin{center}
      $\infer{{{\mbox{\textsf{sp}}}_{X}}\,{\left\{ A\right\} }}{{{\mbox{\textsf{sp}}}_{X}}\,{A}}$
    \end{center}
    }
  \end{displaymath}
  \par
  The fact that $\mathrm{Prop}$ is impredicative -- \emph{i.e.} supports the following product formation rules:
  \begin{displaymath}
    \parbox{1.\textwidth}{
    \begin{center}
      $\infer{\Gamma \vdash \prod _{x{:}A}B : \mathrm{Prop}}{\Gamma \vdash A : s & \Gamma , x{:}A\vdash B : \mathrm{Prop}}$
    \end{center}
    }
  \end{displaymath}
  is easily understood in terms of proof irrelevance. Indeed, if for all $x$, $B$ has at most one element, so has the product over $x$. Even though it uses functional extensionality, which is not provable.
  \par
  \subsection{Singleton rules}
  \par
  The types which (ideally) preserve the proof irrelevance property are sometimes called singleton types in the setting of Coq. In our algebraic presentation, they correspond to inductive type family constructors with extra formation rules to make them preserve propositions. The rules are shown in Figure~\ref{latex_lib_label_5}.
  \begin{figure}[tp]
    \centering{}\makebox[1.\textwidth]{\parbox{0.5\textwidth}{
    \begin{center}
      $\infer{\Gamma \vdash 0 : \mathrm{Prop}}{}$
    \end{center}
    }\parbox{0.5\textwidth}{
    \begin{center}
      $\infer{\Gamma \vdash 1 : \mathrm{Prop}}{}$
    \end{center}
    }}\\
    \makebox[1.\textwidth]{\parbox{1.\textwidth}{
    \begin{center}
      $\infer{\Gamma \vdash {\sum }_{x{:}A}^{f}T : B\rightarrow \mathrm{Prop}}{\Gamma \vdash A : \mathrm{Prop} & \Gamma \vdash B : \mathrm{Prop} & \Gamma , x{:}A\vdash T : \mathrm{Prop} & \Gamma , x{:}A\vdash f : B}$
    \end{center}
    }}\\
    \caption{Singleton rules}\label{latex_lib_label_5}
  \end{figure}
  This elucidates why Coq allows elimination over \textsf{False} and \textsf{And} into arbitrary type: \textsf{False} is implemented as $0$ and \textsf{And}\hphantom{ }\textsf{A}\hphantom{ }\textsf{B} and $A\times B$. The elimination rules being unchanged, pattern-matching over proofs of \textsf{False} and \textsf{And} are unrestricted. Because restricted pattern-matching is often seen as the default, singleton types are said to enjoy \emph{singleton elimination}.
  \par
  Remark that, proofs of propositions being uninformative, there is essentially nothing to be gained from depending on, or being indexed over a proposition. In consequence, the type formation rule for proper indexing in Figure~\ref{latex_lib_label_5} is only useful, in practice, for the subcase of cartesian product.
  \par
  Coq actually implements two other singleton rules. The first one is for inductive fixed points. In our algebraic presentation:
  \begin{displaymath}
    \parbox{1.\textwidth}{
    \begin{center}
      $\infer{\Gamma \vdash \mu {X}^{A\rightarrow \mathrm{Prop}}.\,F : A\rightarrow \mathrm{Prop}}{\Gamma \vdash A : s & \Gamma , X{:}A\rightarrow \mathrm{Prop}\vdash F : A\rightarrow \mathrm{Prop} & {{\mbox{\textsf{sp}}}_{X}}\,{F}}$
    \end{center}
    }
  \end{displaymath}
  It allows to type the accessibility predicate $\mathrm{Acc}$ in $\mathrm{Prop}$. This rule is sound in that fixed points indeed preserve proof irrelevance in presence of functional extensionality. It is also very useful for extraction: structural recursion over $\mathrm{Acc}$ allows the definition of functions whose termination cannot be proved automatically by the guard condition. However, the proof is no longer needed to ensure termination in the target languages of extraction. In this sense, at least, it is static data.
  \par
  The last singleton rule allows properly indexed families in $\mathrm{Prop}$ (not how it is stronger than the rule dependent sum of Figure~\ref{latex_lib_label_5}):
  \begin{displaymath}
    \parbox{1.\textwidth}{
    \begin{center}
      $\infer{\Gamma \vdash {\sum }_{x{:}A}^{f}T : B\rightarrow \mathrm{Prop}}{\Gamma \vdash A : \mathrm{Prop} & \Gamma \vdash B : \mathrm{Type} & \Gamma , x{:}A\vdash T : \mathrm{Prop} & \Gamma , x{:}A\vdash f : B}$
    \end{center}
    }
  \end{displaymath}
  It turns the identity type $\mathrm{Eq}$ into a proposition. It is known to be sound to accept that $\mathrm{Eq}$ is proof irrelevant~\cite{LeeWerner2011}. It is also useful for extraction, as equal types, in a closed environment, are extracted to the same type. Hence a program may safely eliminate over $\mathrm{Eq}$ knowing that it will not affect the performances of the extracted code. In Coq, the index $B$ in the rule above can be of any sort ${\mathrm{Type}}_{i}$, however, this wisdom has been challenged in recent years with the formulation of the \emph{univalence principle}~\cite{hottbook}, of which a simple consequence is that $\mathrm{Eq}$ is \emph{not} proof irrelevant at every type. Indeed, some extracted Coq programs written assuming the univalence principle crash.
  \par
  To correct for the univalence principle, the singleton rule for proper indices can be simply dropped; but it can also be restricted to the lowest sort: $B : {\mathrm{Type}}_{0}$. More precisely the conjunction of the univalence principle and the proof irrelevance principle is consistent as long as the singleton rule of proper indices is restricted to sorts $s$ such that there is no sort $s'$ other than $\mathrm{Prop}$ such that $s' : s$. Because, if such a sort $s'$ exists, $\mathbb{B} : s'$ and by univalence, ${\mathrm{Eq}}\,{\mathbb{B}}$ has two distinct elements contradicting proof irrelevance.
  \par
  For types which do not enjoy singleton elimination, turning them into propositions means restricting their elimination. We achieve this effect by adding a single type construction coercing from $\mathrm{Type}$ to $\mathrm{Prop}$:
  \begin{displaymath}
    \parbox{1.\textwidth}{
    \begin{center}
      \makebox[1.\textwidth]{\parbox{0.5\textwidth}{
      \begin{center}
        $\infer{\Gamma \vdash \left[ A\right]  : \mathrm{Prop}}{\Gamma \vdash A : \mathrm{Type}}$
      \end{center}
      }\parbox{0.5\textwidth}{
      \begin{center}
        $\infer{\Gamma \vdash \left\langle u\right\rangle  : \left[ A\right] }{\Gamma \vdash u : A}$
      \end{center}
      }}\\[0.5mm]
      \makebox[1.\textwidth]{\parbox{1.\textwidth}{
      \begin{center}
        $\infer{\Gamma \vdash 
        \begin{array}{@{}l}
          \mathrm{match}~u~\mathrm{as}~x~\mathrm{return}~P~\mathrm{with}\\
          \begin{array}{|l@{~\Rightarrow ~}l}
            \left\langle y\right\rangle  & v\\
          \end{array}
          \\
        \end{array}
         : {P}{\left[ \,x\,{\setminus }\,u\,\right] }}{\Gamma \vdash u : \left[ A\right]  & \mbox{${\quad}$$\Gamma , x{:}\left[ A\right] \vdash P : \mathrm{Prop}$${\quad}$} & \Gamma , y{:}A\vdash v : {P}{\left[ \,x\,{\setminus }\,\left\langle y\right\rangle \,\right] }}$
      \end{center}
      }}\\[0.5mm]
      \makebox[1.\textwidth]{\parbox{1.\textwidth}{
      \begin{center}
        $
        \begin{array}{@{}l}
          \mathrm{match}~\left\langle u\right\rangle ~\mathrm{as}~x~\mathrm{return}~P~\mathrm{with}\\
          \begin{array}{|l@{~\Rightarrow ~}l}
            \left\langle y\right\rangle  & v\\
          \end{array}
          \\
        \end{array}
        \leadsto {v}{\left[ \,y\,{\setminus }\,u\,\right] }$
      \end{center}
      }}\\[0.5mm]
    \end{center}
    }
  \end{displaymath}
  The important rule is the elimination rule, where the return clause is limited to be of sort $\mathrm{Prop}$, whereas every other type construction can be eliminated to any sort. Apart from this restriction $\left[ A\right] $ is a synonym of $A$, except in $\mathrm{Prop}$. In~\cite{Awodey2004}, the type theory is extensional, in that the identity type and the conversion relation coincide. The elimination rules for bracket is much finer and reflects precisely the fact that propositions are proof-irrelevant. In an intensional type theory, restricting with respect to sorts approximates this behaviour: even if we constrained propositions to be proof-irrelevant, not every proof irrelevant type will have type $\mathrm{Prop}$. The bracketing construction is also strictly positive:
  \begin{displaymath}
    \parbox{1.\textwidth}{
    \begin{center}
      $\infer{{{\mbox{\textsf{sp}}}_{X}}\,{\left[ A\right] }}{{{\mbox{\textsf{sp}}}_{X}}\,{A}}$
    \end{center}
    }
  \end{displaymath}
  \par
  It is actually possible, using only the impredicative dependent product to define a bracketing operation: $\prod _{P{:}\mathrm{Prop}}\left( A\rightarrow P\right) \rightarrow P$. Like $\left[ A\right] $ it behaves as $A$ except it can only be used to form a proposition. However, the impredicative encoding is positive but not strictly, which motivates the introduction of the extra construction.
  \par
  \subsection{Examples}\label{latex_lib_label_8}
  \par
  The logical connectives can be defined as follows:
  \begin{displaymath}
    \left\{ 
    \begin{aligned}
      \mathrm{False}&=0\\
      \mathrm{And}&=\lambda {A}^{\mathrm{Prop}}\,{B}^{\mathrm{Prop}}.\,A\times B\\
      \mathrm{pair}&=\lambda {A}^{\mathrm{Prop}}\,{B}^{\mathrm{Prop}}\,{x}^{A}\,{y}^{B}.\,\left( x,y\right) \\
      \mathrm{Or}&=\lambda {A}^{\mathrm{Prop}}\,{B}^{\mathrm{Prop}}.\,\left[ A+B\right] \\
      {\mathrm{or}}_{\mathrm{introl}}&=\lambda {A}^{\mathrm{Prop}}\,{B}^{\mathrm{Prop}}\,{x}^{A}.\,\left\langle {\mathrm{inl}}\,{x}\right\rangle \\
      {\mathrm{or}}_{\mathrm{intror}}&=\lambda {A}^{\mathrm{Prop}}\,{B}^{\mathrm{Prop}}\,{y}^{B}.\,\left\langle {\mathrm{inr}}\,{y}\right\rangle \\
      \mathrm{Ex}&=\lambda {A}^{\mathrm{Type}}\,{P}^{A\rightarrow \mathrm{Prop}}.\,\left[ \sum _{x{:}A}{P}\,{x}\right] \\
      {\mathrm{ex}}_{\mathrm{intro}}&=\lambda {A}^{\mathrm{Type}}\,{P}^{A\rightarrow \mathrm{Prop}}\,{x}^{A}\,{p}^{{P}\,{x}}.\,\left\langle {\left( x, p\right) }_{{\mbox{\scriptsize{$x${:}$A$.$P$}}}}\right\rangle 
    \end{aligned}
    \right. 
  \end{displaymath}
  Note how, because of the brackets, existentials and disjunctions are prohibited from being eliminated to non-propositional types. Thanks to the singleton rules, however, conjunction and falsity do not require brackets.
  \par
  As a final example, consider the type \textsf{Ascending}\hphantom{ }\textsf{n}\hphantom{ }\textsf{p} of ascending sequences of integers between \textsf{p} and \textsf{n} defined by mutual recursion with the proposition \textsf{Ge}\hphantom{ }\textsf{m}\hphantom{ }\textsf{p} which stands from \textsf{m} is greater than or equal to \textsf{p}:
  \begin{displaymath}
    \parbox{0.8\textwidth}{
    \begin{flushleft}
      \textbf{\textsf{Inductive}}\hphantom{ }\textsf{Ascending}\hphantom{ }\symbol{58}\hphantom{ }\textsf{Nat}\hphantom{ }${\rightarrow}$\hphantom{ }\textsf{Nat}\hphantom{ }${\rightarrow}$\hphantom{ }\textbf{\textsf{Type}}\hphantom{ }\symbol{58}\symbol{61}\\
      ${\mid}$\hphantom{ }\textsf{top}\hphantom{ }\symbol{58}\hphantom{ }\textbf{\textsf{forall}}\hphantom{ }\textsf{n}\symbol{44}\hphantom{ }\textsf{Ascending}\hphantom{ }\textsf{n}\hphantom{ }\textsf{n}\\
      ${\mid}$\hphantom{ }\textsf{up}\hphantom{ }\symbol{58}\hphantom{ }\textbf{\textsf{forall}}\hphantom{ }\textsf{n}\hphantom{ }\textsf{p}\hphantom{ }\textsf{m}\symbol{44}\hphantom{ }\textsf{Ge}\hphantom{ }\textsf{m}\hphantom{ }\symbol{40}\textsf{s}\hphantom{ }\textsf{p}\symbol{41}\hphantom{ }${\rightarrow}$\hphantom{ }\textsf{Ascending}\hphantom{ }\textsf{n}\hphantom{ }\textsf{m}\hphantom{ }${\rightarrow}$\hphantom{ }\textsf{Ascending}\hphantom{ }\textsf{n}\hphantom{ }\textsf{p}\\
      \textbf{\textsf{with}}\hphantom{ }\textsf{Ge}\hphantom{ }\symbol{58}\hphantom{ }\textsf{Nat}\hphantom{ }${\rightarrow}$\hphantom{ }\textsf{Nat}\hphantom{ }${\rightarrow}$\hphantom{ }\textbf{\textsf{Prop}}\hphantom{ }\symbol{58}\symbol{61}\\
      ${\mid}$\hphantom{ }\textsf{ascend}\hphantom{ }\symbol{58}\hphantom{ }\textbf{\textsf{forall}}\hphantom{ }\textsf{m}\hphantom{ }\textsf{p}\symbol{44}\hphantom{ }\textsf{Ascending}\hphantom{ }\textsf{m}\hphantom{ }\textsf{p}\hphantom{ }${\rightarrow}$\hphantom{ }\textsf{Ge}\hphantom{ }\textsf{m}\hphantom{ }\textsf{p}\symbol{46}
    \end{flushleft}
    }
  \end{displaymath}
  As \textsf{Ascending} has type \textbf{\textsf{Type}}, whereas \textsf{Ge} has type \textbf{\textsf{Prop}}, the translation to a single inductive type is not as straightforward as \textsf{Even} and \textsf{Odd}. The translation requires the use of brackets around the recursive calls:
  \begin{displaymath}
    \left\{ 
    \begin{aligned}
      \mathrm{AscendingGe}&=
      \begin{array}{l}
        \mu {X}^{\left( \mathrm{Nat}\times \mathrm{Nat}\right) +\left( \mathrm{Nat}\times \mathrm{Nat}\right) \rightarrow \mathrm{Type}}.\,\lambda {i}^{\left( \mathrm{Nat}\times \mathrm{Nat}\right) +\left( \mathrm{Nat}\times \mathrm{Nat}\right) }.\,\mbox{\phantom{a}}\\
        \begin{array}{cl}
           & {\left( {\sum }_{n{:}\mathrm{Nat}}^{{\mathrm{inl}}\,{\left( n,n\right) }}1\right) }\,{i}\\
          + & {\left( {\sum }_{j{:}\mathrm{Nat}\times \mathrm{Nat}}^{{\mathrm{inl}}\,{j}}\sum _{m{:}\mathrm{Nat}}\left[ {X}\,{\left( {\mathrm{inr}}\,{\left( m,{s}\,{\left( {{\pi }_{2}}\,{j}\right) }\right) }\right) }\right] \times {X}\,{\left( {\mathrm{inl}}\,{\left( {{\pi }_{1}}\,{j},m\right) }\right) }\right) }\,{i}\\
          + & {\left( {\sum }_{j{:}\mathrm{Nat}\times \mathrm{Nat}}^{{\mathrm{inr}}\,{j}}{X}\,{\left( {\mathrm{inl}}\,{j}\right) }\right) }\,{i}\\
        \end{array}
        \\
      \end{array}
      \\
      \mathrm{Ascending}&=\lambda {n}\,{p}.\,{\mathrm{AscendingGe}}\,{\left( {\mathrm{inl}}\,{\left( n,p\right) }\right) }\\
      \mathrm{Ge}&=\lambda {m}\,{p}.\,\left[ {\mathrm{AscendingGe}}\,{\left( {\mathrm{inr}}\,{\left( m,p\right) }\right) }\right] 
    \end{aligned}
    \right. 
  \end{displaymath}
  \section{Impredicative Set}
  \par
  In addition to the impredicative sort $\mathrm{Prop}$, Coq has a sort $\mathrm{Set}$ which is predicative by default but can be turned impredicative with a flag. Where $\mathrm{Prop}$ is meant to be used in the context of separating static and dynamic information, the spirit of the impredicative sort $\mathrm{Set}$ is to be as powerful as possible without being inconsistent. In the algebraic presentation, that means being stable by every construction except dependent sums with the first projection in an arbitrary sort (\emph{strong sums}).
  \par
  To mirror the optional nature of the impredicativity of $\mathrm{Set}$, the rules for a predicative sort $\mathrm{Set}$ are given in Figure~\ref{latex_lib_label_6}; to turn impredicativity on, the rules of Figure~\ref{latex_lib_label_7} must be used \emph{in addition} to those of predicative $\mathrm{Set}$. This presentation makes immediately apparent that impredicative $\mathrm{Set}$ is an extension of predicative $\mathrm{Set}$, in that every program of the latter typechecks in the former.
  \par
  The rules of $\mathrm{Set}$ are the same as those of $\mathrm{Prop}$, with the exception of $A+B$ which is in $\mathrm{Set}$ when both $A$ and $B$ are -- even with predicative $\mathrm{Set}$. Hence, there are types in $\mathrm{Set}$ with several elements -- \emph{e.g.} $\mathbb{B}$. As a consequence, the bracketing operation which coerces types in $\mathrm{Type}$ to $\mathrm{Set}$ does not enjoy an explanation in terms of proof irrelevance, as was the case in $\mathrm{Prop}$. As a matter of fact, there is no clear set-theoretical description at all. A close cousin of $\mathrm{Set}$ bracketing, however, can be found in homotopy type theory~\cite{hottbook}, where, roughly, groupoids are \emph{truncated} to sets through a quotient of their homsets by the total relation.
  \par
  \begin{figure}[tp]
    \centering{}\makebox[1.\textwidth]{\parbox{0.5\textwidth}{
    \begin{center}
      $\infer{\Gamma \vdash \mathrm{Set} : \mathrm{Type}}{}$
    \end{center}
    }\parbox{0.5\textwidth}{
    \begin{center}
      $\infer{\Gamma \vdash \prod _{x{:}A}B : \mathrm{Set}}{\Gamma \vdash A : \mathrm{Set} & \Gamma , x{:}A\vdash B : \mathrm{Set}}$
    \end{center}
    }}\\[0.5mm]
    \makebox[1.\textwidth]{\parbox{1.\textwidth}{
    \begin{center}
      $\infer{\Gamma \vdash \sum _{x{:}A}P : \mathrm{Set}}{\Gamma \vdash A : \mathrm{Set} & \Gamma , x{:}A\vdash P : \mathrm{Set}}$
    \end{center}
    }}\\[0.5mm]
    \makebox[1.\textwidth]{\parbox{0.333333333333\textwidth}{
    \begin{center}
      $\infer{\Gamma \vdash A+B : \mathrm{Set}}{\Gamma \vdash A : \mathrm{Set} & \Gamma \vdash B : \mathrm{Set}}$
    \end{center}
    }\parbox{0.333333333333\textwidth}{
    \begin{center}
      $\infer{\Gamma \vdash 1 : \mathrm{Set}}{}$
    \end{center}
    }\parbox{0.333333333333\textwidth}{
    \begin{center}
      $\infer{\Gamma \vdash 0 : \mathrm{Set}}{}$
    \end{center}
    }}\\[0.5mm]
    \makebox[1.\textwidth]{\parbox{1.\textwidth}{
    \begin{center}
      $\infer{\Gamma \vdash \mu {X}^{A\rightarrow \mathrm{Set}}.\,F : A\rightarrow \mathrm{Set}}{\Gamma \vdash A : s & \Gamma , X{:}A\rightarrow \mathrm{Set}\vdash F : A\rightarrow \mathrm{Set} & {{\mbox{\textsf{sp}}}_{X}}\,{F}}$
    \end{center}
    }}\\[0.5mm]
    \makebox[1.\textwidth]{\parbox{1.\textwidth}{
    \begin{center}
      $\infer{\Gamma \vdash {\sum }_{x{:}A}^{f}T : B\rightarrow \mathrm{Set}}{\Gamma \vdash A : \mathrm{Set} & \Gamma \vdash B : \mathrm{Set} & \Gamma , x{:}A\vdash T : \mathrm{Set} & \Gamma , x{:}A\vdash f : B}$
    \end{center}
    }}\\[0.5mm]
    \makebox[1.\textwidth]{\parbox{1.\textwidth}{
    \begin{center}
      $\infer{\Gamma \vdash \nu {X}^{A\rightarrow \mathrm{Set}}.\,F : A\rightarrow \mathrm{Set}}{\Gamma \vdash A : s & \Gamma , X{:}A\rightarrow \mathrm{Set}\vdash F : A\rightarrow \mathrm{Set} & {{\mbox{\textsf{sp}}}_{X}}\,{F}}$
    \end{center}
    }}\\[0.5mm]
    \makebox[1.\textwidth]{\parbox{0.5\textwidth}{
    \begin{center}
      $\infer{\Gamma \vdash {\left\{ A\right\} }_{\mathrm{Set}} : \mathrm{Type}}{\Gamma \vdash A : \mathrm{Set}}$
    \end{center}
    }\parbox{0.5\textwidth}{
    \begin{center}
      $\infer{\Gamma \vdash {\mathrm{elt}}\,{u} : {\left\{ A\right\} }_{\mathrm{Set}}}{\Gamma \vdash u : A}$
    \end{center}
    }}\\[0.5mm]
    \makebox[1.\textwidth]{\parbox{1.\textwidth}{
    \begin{center}
      $\infer{\Gamma \vdash 
      \begin{array}{@{}l}
        \mathrm{match}~u~\mathrm{as}~x~\mathrm{return}~P~\mathrm{with}\\
        \begin{array}{|l@{~\Rightarrow ~}l}
          {\mathrm{elt}}\,{y} & v\\
        \end{array}
        \\
      \end{array}
       : {P}{\left[ \,x\,{\setminus }\,u\,\right] }}{\Gamma \vdash u : {\left\{ A\right\} }_{\mathrm{Set}} & \Gamma , x{:}{\left\{ A\right\} }_{\mathrm{Set}}\vdash P : s & \Gamma , y{:}A\vdash v : {P}{\left[ \,x\,{\setminus }\,{\mathrm{elt}}\,{y}\,\right] }}$
    \end{center}
    }}\\[0.5mm]
    \makebox[1.\textwidth]{\parbox{1.\textwidth}{
    \begin{center}
      $
      \begin{array}{@{}l}
        \mathrm{match}~{\mathrm{elt}}\,{u}~\mathrm{as}~x~\mathrm{return}~P~\mathrm{with}\\
        \begin{array}{|l@{~\Rightarrow ~}l}
          {\mathrm{elt}}\,{y} & v\\
        \end{array}
        \\
      \end{array}
      \leadsto {v}{\left[ \,y\,{\setminus }\,u\,\right] }$
    \end{center}
    }}\\[0.5mm]
    \makebox[1.\textwidth]{\parbox{1.\textwidth}{
    \begin{center}
      $\infer{{{\mbox{\textsf{sp}}}_{X}}\,{{\left\{ A\right\} }_{\mathrm{Set}}}}{{{\mbox{\textsf{sp}}}_{X}}\,{A}}$
    \end{center}
    }}\\[0.5mm]
    \caption{Rules for predicative $\mathrm{Set}$}\label{latex_lib_label_6}
  \end{figure}
  \begin{figure}[tp]
    \centering{}\makebox[1.\textwidth]{\parbox{0.5\textwidth}{
    \begin{center}
      $\infer{\Gamma \vdash \prod _{x{:}A}B : \mathrm{Set}}{\Gamma \vdash A : s & \Gamma , x{:}A\vdash B : \mathrm{Set}}$
    \end{center}
    }\parbox{0.5\textwidth}{
    \begin{center}
      $\infer{{{\mbox{\textsf{sp}}}_{X}}\,{{\left[ A\right] }_{\mathrm{Set}}}}{{{\mbox{\textsf{sp}}}_{X}}\,{A}}$
    \end{center}
    }}\\[0.5mm]
    \makebox[1.\textwidth]{\parbox{0.5\textwidth}{
    \begin{center}
      $\infer{\Gamma \vdash {\left[ A\right] }_{\mathrm{Set}} : \mathrm{Set}}{\Gamma \vdash A : \mathrm{Type}}$
    \end{center}
    }\parbox{0.5\textwidth}{
    \begin{center}
      $\infer{\Gamma \vdash {\left\langle u\right\rangle }_{\mathrm{Set}} : {\left[ A\right] }_{\mathrm{Set}}}{\Gamma \vdash u : A}$
    \end{center}
    }}\\[0.5mm]
    \makebox[1.\textwidth]{\parbox{1.\textwidth}{
    \begin{center}
      $\infer{\Gamma \vdash 
      \begin{array}{@{}l}
        \mathrm{match}~u~\mathrm{as}~x~\mathrm{return}~P~\mathrm{with}\\
        \begin{array}{|l@{~\Rightarrow ~}l}
          {\left\langle y\right\rangle }_{\mathrm{Set}} & v\\
        \end{array}
        \\
      \end{array}
       : {P}{\left[ \,x\,{\setminus }\,u\,\right] }}{\Gamma \vdash u : {\left[ A\right] }_{\mathrm{Set}} & \Gamma , x{:}{\left[ A\right] }_{\mathrm{Set}}\vdash P : \mathrm{Set} & \Gamma , y{:}A\vdash v : {P}{\left[ \,x\,{\setminus }\,{\left\langle y\right\rangle }_{\mathrm{Set}}\,\right] }}$
    \end{center}
    }}\\[0.5mm]
    \makebox[1.\textwidth]{\parbox{1.\textwidth}{
    \begin{center}
      $
      \begin{array}{@{}l}
        \mathrm{match}~{\left\langle u\right\rangle }_{\mathrm{Set}}~\mathrm{as}~x~\mathrm{return}~P~\mathrm{with}\\
        \begin{array}{|l@{~\Rightarrow ~}l}
          {\left\langle y\right\rangle }_{\mathrm{Set}} & v\\
        \end{array}
        \\
      \end{array}
      \leadsto {v}{\left[ \,y\,{\setminus }\,u\,\right] }$
    \end{center}
    }}\\[0.5mm]
    \caption{Rules for impredicative $\mathrm{Set}$}\label{latex_lib_label_7}
  \end{figure}
  \par
  \section{Conclusion}
  \par
  The algebraic presentation of Coq makes the conversion between sorts explicit. The toplevel inductive definitions of Coq can be understood as implicitly inserting canonical bracketing operations when an inductive type is declared inside an impredicative sort but should be of a different sort due to its form; and inserting type coercion from a smaller sort to a bigger sort when applying a cumulativity rule.
  \par
  Monolithic type definitions like in Coq have a number of advantages over the algebraic presentation, they boil down to better type errors due to naming, better type inference and better memory representation due to $n$-ary sums and products. However, the value of the implicit coercions between sorts is less clear. In particular, the bracketing operation to impredicative sorts is probably a better guide for program extraction than the current method of figuring whether or not a given type is a proposition, which interacts badly with universe polymorphism~\cite{Letouzey2005}. Explicit coercions for extraction are also in the spirit of~\cite{Barras2008}.
  \par
  All of the algebraic type constructors can actually be defined in Coq, except the two fixed-points because there is no way to abstract over strictly positive type families. So is it clear that expressions of the algebraic presentation which do not use inductive or co-inductive fixed points can be translated into Coq. Occurrences of the fixed points in a type must be ${\lambda}$-lifted and given a toplevel name. Some care must be given to avoiding the duplication of such definitions otherwise types which must be convertible for the expression to typecheck, might be seen as different in the Coq translation. Apart from this technicality, translation from the algebraic presentation to Coq is straightforward. We claim that, at least if we extend the algebraic presentation to a hierarchy of universes and the strict positivity condition is made a bit more fine-grained, Coq terms can be, conversely, translated into the algebraic presentation.
  \bibliography{library}
\end{document}